\documentclass[twocolumn,showpacs,preprintnumbers,amsmath,amssymb,superscriptaddress,pre]{revtex4}

\usepackage{graphicx}% Include figure files
\usepackage{dcolumn}% Align table columns on decimal point
\usepackage{bm}% bold math

\begin{document}

\title{The totally asymmetric exclusion process with extended objects, a model for protein synthesis}

\author{Leah B.\ Shaw}
 \email{lbs22@cornell.edu}
 \affiliation{Department of Physics, Cornell University, Ithaca, NY 14853-2501}
 \affiliation{School of Chemical and Biomolecular Engineering,
 Cornell University, Ithaca, NY 14853-5201}

\author{R.K.P.\ Zia}
 \affiliation{Center for Stochastic Processes in Science and Engineering, Physics Department,\\
 Virginia Polytechnic Institute and State University, Blacksburg, VA 24061-0435}

\author{Kelvin H.\ Lee}
 \affiliation{School of Chemical and Biomolecular Engineering,
 Cornell University, Ithaca, NY 14853-5201}

\date{\today}

\begin{abstract}
The process of protein synthesis in biological systems resembles a
one dimensional driven lattice gas in which the particles have
spatial extent, covering more than one lattice site.  We expand
the well studied Totally Asymmetric Exclusion Process (TASEP), in
which particles typically cover a single lattice site, to include
cases with extended objects.  Exact solutions can be determined
for a uniform closed system.  We analyze the uniform open system
through two approaches. First, a continuum limit produces a
modified diffusion equation for particle density profiles. Second,
an extremal principle based on domain wall theory accurately
predicts the phase diagram and currents in each phase.  Finally,
we briefly consider approximate approaches to a non-uniform open
system with quenched disorder in the particle hopping rates and
compare these approaches with Monte Carlo simulations.
\end{abstract}


\maketitle

\section*{Introduction}

The process of protein synthesis, important in biological systems,
has been the focus of intensive study over the last few decades.
We concentrate on protein synthesis in prokaryotes, particularly
\textit{Escherichia coli}, because it is relatively simple and
well studied. The mechanism consists of ribosomes ``reading'' the
codons of messenger RNA (mRNA) as the ribosomes move along an mRNA
chain, and the recruitment and assembly of amino acids
(appropriate to the codons being read) to form a protein. (See,
e.g., \cite {Stryer}, for more details.) Known as ``translation,''
this process is often described as three steps: initiation, where
ribosomes attach themselves, one at a time, at the ``start'' end
of the mRNA; elongation, where the ribosomes move down the chain
in a series of steps; and termination, where they detach at the
``stop'' codon. Since ribosomes cannot overlap, their dynamics is
subject to the ``excluded volume constraint.'' Apart from being
impeded by another ribosome (steric hinderance), a ribosome cannot
move until the arrival of an appropriate transfer RNA, carrying
the appropriate amino acid (a combination known as aminoacyl-tRNA,
or aa-tRNA). Thus, the relative abundances of the approximately 60
types \cite{NU} of aa-tRNA have significant effects on the
elongation rate. Assuming reactant availabilities in a cell are in
their steady state, with a time-independent concentration of
ribosomes and aa-tRNA, there would be an approximately steady
(average) current of ribosomes moving along the mRNA, resulting in
a specific production rate of this particular protein. Our goal is
the prediction of the protein production rates for various mRNA's,
as a function of the concentration of ribosomes and aa-tRNA's.

The process of translation is well suited to modeling using a
driven lattice gas in one dimension. In most relevant studies of
one dimensional driven lattice gases, particles are injected at
some rate on one end of a chain of discrete lattice sites, then
hop down the chain one site at a time with another rate, and
finally exit the chain at the other end with a third rate. These
three rates correspond to the rates of initiation, elongation, and
termination. The excluded volume constraint is implemented by
insuring that each site can accommodate at most one particle.
Because the dynamics is stochastic, even this ``simple'' model,
though soluable exactly \cite{DDM, SD}, is already not trivial.
However, we believe two essential aspects of translation are
missing from this model. First, if we model a codon by a lattice
site, the ribosome would cover, typically, a dozen sites
\cite{Heinrich, Kang}. Second, there is non-uniformity in the
hopping (elongation) rates along the chain, because a ribosome has
to ``wait'' for the appropriate aa-tRNA before continuing, and the
relative abundance of the different aa-tRNA's is far from unity.
Remarkably, the first issue was explored as early as 1968
\cite{MGP, MG}, though only at the deterministic, ``mean field''
level.

In this paper, we present studies which address both of these
issues, extending the work on the ``simple model'' known as TASEP
(namely, single-site coverage, Totally Asymmetric Simple Exclusion
Processes with open boundaries) \cite{GS-DL19}. Our methods
involve both Monte Carlo simulations and modern analysis
techniques, including domain wall theory \cite{KSKS}. We have
confirmed all key results from earlier studies \cite {MGP, MG},
and several new insights have emerged. The paper is organized as
follows. The details of the model are delineated first, along with
brief summaries of known results. Section \ref{sec:ring} is
devoted to a \emph{closed} (i.e., periodic) system with particles
of arbitrary size. Though not a direct model for translation,
TASEP on a ``ring'' provides simple solutions as well as useful
insights in the form of \emph{exact} relations for relevant
parameters, such as current-density relations. In Section
\ref{sec:open}, we turn to the central topic: TASEP with extended
objects and open boundaries. Section \ref{sec:disorder} is devoted
to non-uniform hopping (elongation) rates. We close with a brief
summary and speculate on the relevance of our model as a mechanism
for the nonlinear relationship between mRNA and protein levels
observed in biological experiments \cite{Gygi, Ideker}.

\section{The driven lattice gas as a model for protein synthesis}
\label{sec:model}

\subsection{Model specifications}

We model an mRNA with $N$ codons as a chain of sites, each of
which is labeled by $i$. The first and last sites, $i=1,N$, are
associated with the start and stop codons respectively. At any
time, attached to the mRNA are $M$
ribosomes (also referred to as ``particles''), which we label by $\alpha $ ($
\alpha =1,...,M$). Being a large complex of molecules, each
ribosome will cover $\ell $ sites (codons), with $\ell =12$
typically \cite{Heinrich, Kang}. By contrast, nearly all studies
of the asymmetric simple exclusion processes (ASEP) are
devoted to $\ell =1$. Any site may be covered by a single ribosome
or none. In case of the latter, we will refer to the site as
``empty'' or ``occupied by a hole.'' For convenience, we define
$\tilde{M}$ as the number of holes on the chain, so that
\begin{equation}
\tilde{M}+\ell M=N. \label{MMN}
\end{equation}
For open systems, a ribosome at the end can be attached without
covering all $\ell $ codons, so that this equality is only
approximately true. To locate the ribosome, we arbitrarily choose
the \emph{lowest} site covered. For example, if the first $\ell
\,$ sites are empty, a ribosome can bind in an initiation step,
and then it is said to be ``on site $i=1.$'' Therefore, a complete
specification of the configuration (state, or micro-state) of the
mRNA is the set of locations: $\left\{ i_{\alpha }\right\} $. One
disadvantage of this labeling is that, with each initiation event,
the $\alpha $ of every ribosome will change (increase by unity).
Alternatively, we can use \emph{site occupation} numbers
\begin{equation*}
n_{i}=\left\{
\begin{tabular}{ll}
$1$ & $\text{if site }i\text{ is covered by any part of a ribosome}$ \\
$0$ & $\text{if site }i\text{ is empty}\,.$
\end{tabular}
\right.
\end{equation*}

With these conventions, we define several density parameters:
\begin{itemize}
\item $\rho _r\equiv M/N$ is the ribosome (or particle) density;
\item $\rho \equiv M\ell /N=\sum_in_i/N$ is the coverage density;
\item $\rho _h=1-\rho $ is the hole density; and

\item $\rho _{s}\equiv \rho _{r}+\rho _{h}$ is defined for
convenience.
\end{itemize}
All of these quantities are time dependent, because initiation and
termination occur \emph{independently}. For mathematical reasons,
we will first consider a \emph{closed} system (with periodic
boundary conditions, i.e., the ends of the chain tied to form a
ring), for which these densities are fixed and Eqn.\ (\ref{MMN})
holds strictly. As will be clear later, it is also convenient to
label configurations by specifying the number of holes between
successive ribosomes: $\left\{ h_{\alpha }\right\} $, where
$h_{\alpha }$ is the number of holes \emph{in front of} the
$\alpha ^{th}$ ribosome. In the terminology of traffic models, $
h_{\alpha }$ is also known as the ``headway'' of this particle.
Though not absolutely necessary, we could define $h_{0}$ as the
number of holes behind the first ribosome.

Next, we specify the dynamics of our model. An attached ribosome
located at site $i$ will move to the next site ($i+1$) with a rate
$k_{i}$, \emph{provided} site $i+\ell $ is empty. For Monte Carlo
simulations, it is convenient to update configurations in discrete
time units. Then, it is better to use probabilities $p_{i}$
($0\leq p_{i}\leq 1$), so that a ribosome on site $i$ will be
moved or not with probability $p_{i}$ or $ 1-p_{i}$, respectively.
We purposefully associate these hopping probabilities with a site
because a site is associated with a particular codon. Thus, the
jump rate from that site may depend on the relative abundance of
the appropriate aa-tRNA. Apart from these probabilities, another
aspect of our stochastics is random sequential updating: i.e.,
during each Monte Carlo step (MCS), $M+1$ particles are chosen at
random, in sequence, to attempt moves. They are selected from a
pool that includes the $M$ particles on the lattice plus another
unbound particle that can initiate if there are $\ell $ holes at
the beginning of the chain. Let us illustrate with a few examples.
First, $p_{0}$ is associated with the start codon and, \emph{if
}the first $\ell $ sites are empty, a particle will be placed on
the $i=1$ site with this probability. Next, a random particle (say
at site $i$) is chosen and, provided it has a hole in front ($
n_{i+\ell }=0$), will be moved with probability $p_{i}$.
Naturally, it will not be necessary to check for headway for the
``last'' ribosome ($\alpha =M$ ). Finally, the stop codon will be
associated with $p_{N}$. For simulations of the closed system,
there will be no ``beginning'' or ``end,'' so that there are no
special steps for initiation or termination.

In our computational studies, 100 identical systems of $N$ sites
are simulated in parallel to obtain good statistics. Simulations
of closed systems begin with particles evenly distributed around
the ring and run for 3600 MCS to ensure that steady state is
reached. Open systems begin empty and are run for 12,000 MCS (for
$N<500$) or $100N$ MCS (for $N\geq 500$) to reach steady state.
After steady state is attained, data including the current and
density distribution can be collected.  Density data is typically
collected every 100 MCS. We often use continuous time Monte Carlo
\cite{BKL} because it runs far more quickly than and provides the
same results as standard Monte Carlo.

\subsection{Brief survey of known results}

Extensive investigations of the simple totally asymmetric
exclusion process (TASEP, defined as point particles hopping with
unit rate along a line) with open boundaries can be found in the
literature. Simulations have been performed \cite{Krug}, and exact
analytic results for the steady state exist \cite{DDM, SD}.
Depending on the initiation (or injection) and termination (or
depletion) rates, the system will settle into one of three phases.
Introduced above as $p_0$ and $p_N$ respectively, the initiation
and termination probabilities are mostly referred to as simply
$\alpha $ and $\beta $ in the literature. From their dominant
characteristics, the three phases are known as low density, high
density, and maximal current. A phase diagram in this $\alpha
$-$\beta $ plane has been determined, showing second order
transitions between the maximal current phase and the others, as
well as a first order transition between the high- and low-density
regions. Subtle correlations further divide the last two into
subregions. When disorder is introduced, i.e., not all the $
p_i$'s are equal, then methods for exact analytic approaches fail
(except in the extremely dilute limit, where only the motion of a
single particle is of concern \cite{Derrida}). Indeed, even a
single slow rate in a \emph{closed} system poses serious
difficulties \cite{JL1, JL2, Sep}. However, Kolomeisky \cite{ABK}
obtained approximate steady state solutions and phase diagrams for
an \emph{open} system with a single nonuniform rate in the bulk by
splitting the system into two smaller systems connected by the
nonuniform rate. Tripathy and Barma \cite{Barma} considered a
closed system, but with a finite fraction of identical slow sites.
Based on a combination of Monte Carlo simulations and numerical
solutions of mean field equations, they found current-density
relations. Though not a model for elongation, a related problem is
``particlewise disorder,'' in which the unequal hopping rates are
associated with the particles rather than the sites
\cite{Bengrine}. Further references on TASEP with disorder may be
found in a recent review \cite {JK-BJP}. Also indirectly related
to our one dimensional models are driven lattice gases with
quenched disorder in \emph{higher} dimensions \cite {LF, BJ}.
Finally, we mention that there are many studies on the ASEP in
which a particle has a finite probability of stepping backwards
\cite{ASEP}. Back steps are not generally believed to occur in
elongation, and we will not consider such processes. All of these
studies are restricted to $\ell =1 $.

Systems with extended objects ($\ell >1$) have been rarely
investigated, despite their introduction over three decades ago as
a model for biopolymerization \cite{MGP}. Using a mean field
approach, MacDonald \textit{et al.} set up mean field equations
for the average site occupation $ \left\langle n_i\right\rangle $
and considered both closed \cite{MGP} and open \cite{MG} systems.
In the former case, exact solutions were found, leading them to a
current vs. density relation. For the latter, the authors resorted
to numerical solutions to find the phase diagram for a variety of
initiation and termination rates. A phase diagram similar to the
simple TASEP, as well as non-trivial density profiles and the
associated currents, was obtained. More recently, there is renewed
interest in this problem. Naming this system ``$\ell $-TASEP,''
Sasamoto and Wadati \cite{SW} focused on the time dependence of
$M$ particles in an infinite lattice and, using the Bethe ansatz,
found exact results for the conditional probability that the
particles are found at certain sites given an initial
configuration. The main conclusion is that the dynamics of $\ell
$-TASEP lies in the same universality class as ordinary TASEP.
This line of inquiry has been further generalized to a system
containing a \emph{distribution }of particle sizes \cite{AB, FA}.
Though these investigations produced interesting results, they are
not applicable to our situation, namely, finite systems with open
boundaries. In particular, for finite lattices, these studies are
restricted to \emph{closed} systems, for which the stationary
states are trivial, as we will recapitulate in Section
\ref{sec:ring}. Finally, a recent work by Lakatos and Chou
\cite{LC} considered uniform open systems with extended objects.
Using a discrete Tonks gas partition function, they derived the
current vs. density relation first presented by MacDonald
\textit{et al.} \cite{MGP}. Via a refined mean field theory, they
extended this result to predict currents and bulk densities for
the open system, which they confirmed by Monte Carlo simulations.
The phase diagram and its properties are consistent with those
initially obtained by MacDonald and Gibbs \cite{MG}. We are not
aware of published results on open systems with both extended
objects and quenched disorder.

\section{TASEP of extended objects on a ring}
\label{sec:ring}

For simplicity and mathematical reasons, it is convenient to
discuss a uniform closed system with periodic boundary conditions. Here,
$M$ identical particles of size $\ell $ move on a ring of $N$
sites. Due to the excluded volume constraint, there are
$\tilde{M}=N-\ell M$ uncovered sites (holes). The system evolves
by discrete time steps, with random sequential updates.
Specifically, during each MCS, $M$ particles are chosen at random
in sequence, and each is moved forward one site provided there is
headway. Since
we have a stochastic dynamics, the complete description of this system is $
P\left( C,t\right) $, the probability that it is found in
configuration $C$ after $t$ steps (starting from some initial
$C_{0}$). To label a configuration, we choose an arbitrary
particle to be the first ($\alpha =1$)
and supply $\left\{ h_{\alpha }\right\} $, the set of number of holes in front of the $
\alpha ^{th}$ particle (with $\alpha =1,...,M$). Clearly, we may
also think
of a configuration as a series of $M$ ``gaps'' (between the particles) with $
h_{\alpha }$ being the number of holes in the $\alpha ^{th}$ gap. So, $
P\left( h_{1},h_{2},...h_{M};t\right) $ is an explicit form for
$P\left( C,t\right) $. Note that, since the system is closed,
there is a constraint on $h_{\alpha }$, i.e.,
\begin{equation*}
\sum_{\alpha }h_{\alpha }=\tilde{M}  \label{M-tilde}
\end{equation*}
is a constant in time.

Random sequential updating can be translated into an equation
which governs the time evolution of $P\left( C,t\right) $, namely
a master equation. Starting with the initial $P\left( C,0\right)
=\delta \left( C,C_0\right) $ (where $\delta $ is the Kronecker
delta), $P\left( C,t\right) $ is expected to settle into a unique,
\emph{time-independent} distribution, $P^{*}\left( C\right) $,
which we will refer to as the ``steady state.'' If this system
were evolving towards thermal equilibrium, the dynamics would
satisfy detailed balance \cite{DB}, and $P^{*}\left( C\right) $
would be given by the well-known Boltzmann factor. However, the
dynamics of our system definitely violates detailed balance, so
that, associated with a \emph{non-equilibrium} steady state,
$P^{*}\left( C\right) $ is not known in general. Fortunately, this
closed system belongs to a class for which a simple solution is
known \cite{Spitzer}; namely, every configuration occurs with
equal probability \cite{pairwisenote}. Thus, $P^{*}\left( C\right)
$ is precisely the reciprocal of the total number of
configurations consistent with the given parameters ($N,M,\ell $).
With such a simple distribution, we can compute many quantities of
interest, such as the probability distribution of the current (and
hence the average current) and the headway.

Apart from overall factors, the total number of configurations is
just $Z(\tilde{M},M),$ the total number of lists $\left\{ h_\alpha
\right\} $ subject to Eqn.\ (\ref{M-tilde}). Because this is a
well-known combinatorial problem (appearing in, e.g., the Bose
gas), we simply quote the result:
\begin{equation*}
Z(\tilde{M},M)=\frac{(\tilde{M}+M-1)!}{\tilde{M}!(M-1)!}\equiv
\binom{\tilde{ M}+M-1}{\tilde{M}}
\end{equation*}
The actual number of distinct configurations is $\left( N/M\right)
Z(\tilde{M },M),$ because there are $N$ lattice sites on which the
first particle can be placed but the $M$ particles are identical.
Thus,
\begin{equation*} P^{*}\left( C\right)
=\frac{\tilde{M}!M!}{N(\tilde{M}+M-1)!}
\end{equation*}
\emph{independent} of $C$.

\begin{figure}[tbp]
\includegraphics[clip=true]{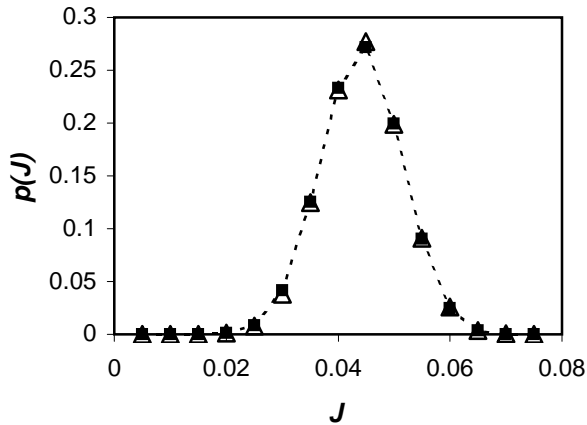}
\caption{Distribution of currents $p(J)$ for $N=200$, $M=15$, and
$\ell =12$. Squares (connected by dashed lines) are values
predicted by Eqn.\ (\ref {PofJfromM}), and triangles are values
observed in Monte Carlo simulations. A single lattice was
simulated to steady state, and instantaneous current ($ H/N $) was
determined every 100 MCS thereafter for $1.2\times 10^{6}$ MCS.}
\label{fig:curDist}
\end{figure}

To compute the probability distribution of the current and its
average, we need more detailed information on the above
``partition.'' Defining the current, $J$, as the number of
particles which moved in one step (normalized by the system size
$N$), we see that we will need $H$, the number of gaps with one or
more holes. The reason is that, for each such gap, the ribosome
behind it can move and contributes one ``unit'' to the current, so
that
\begin{equation*}
J=\frac{H}{N}.
\end{equation*}
Since all configurations are equally probable, the statistical
weight associated with this $J$ is just the total number of
configurations with a
given $H$. This quantity, denoted by $Z(H;\tilde{M},M)$ in analog to $Z(
\tilde{M},M)$, may be found from its definition:

\begin{eqnarray*}
Z(H;\tilde{M},M)&=&\sum_{\left\{ h_{\alpha }\right\} } \left[
 \delta \left( \tilde{M} ,\sum_{\alpha }h_{\alpha }\right) \times
 \right. \nonumber
 \\
& & \left. \delta \left( H,\sum_{\alpha }\left[ 1-\delta \left(
h_{\alpha },0\right) \right] \right) \right] ,
\end{eqnarray*}
where the sum is over all possible lists $\left\{ h_{\alpha
}\right\} $ and the Kronecker $\delta $'s select only those which
satisfy Eqn.\ (\ref{M-tilde}) and have $H$ gaps with $h_{\alpha
}>0$. Once $Z(H;\tilde{M},M)$ is known, the full distribution for
the current is
\begin{equation*}
p(J;\tilde{M},M)=\frac{Z(H;\tilde{M},M)}{Z(\tilde{M},M)}.
\end{equation*}
Now, the explicit form of $Z(H;\tilde{M},M))$ can be obtained
either through the generating function
\begin{eqnarray*}
W_{M}(\zeta ,\eta )&\equiv&
\sum_{\tilde{M},H}Z(H;\tilde{M},M)\zeta ^{\tilde{M}
}\eta ^{H} \nonumber \\
&=&\left[ 1+\frac{\zeta \eta }{1-\zeta }\right] ^{M}
\end{eqnarray*}
or by standard combinatorial techniques:
\begin{equation*}
Z(H;\tilde{M},M)=\binom{M}{H}\binom{\tilde{M}-1}{H-1}.
\end{equation*}
Thus, the explicit current distribution is
\begin{equation}
p(J)=\binom{M}{H}\binom{\tilde{M}-1}{H-1}\left/ \binom{\tilde{M}+M-1}{\tilde{
M}}\right. \,,  \label{PofJfromM}
\end{equation}
where $H$ on the right stands for $JN.$ An alternate form, showing
the dependence of this distribution on the control parameters
($\tilde{M},M,N= \tilde{M}+\ell M$) is

\begin{eqnarray*}
\textstyle p(J\,|\,\tilde{M},M,N) &=& \frac{1}{M\tilde{M}\left(
\tilde{M}+M-1\right) !} \times
 \\
& &
\frac{JN}{(\tilde{M}-JN)!(M-JN)!} \times
 \\
& &
\left[ \frac{M!\tilde{M}!}{\left(
JN\right) !}\right] ^{2}.
\end{eqnarray*}
To illustrate this distribution, we show in Figure
\ref{fig:curDist} both this prediction and simulation data for the
case of $N=200$, $M=15$, and $\ell =12$. Clearly, there is
excellent agreement between theory and simulation.

\begin{figure}[tbp]
\includegraphics[clip=true]{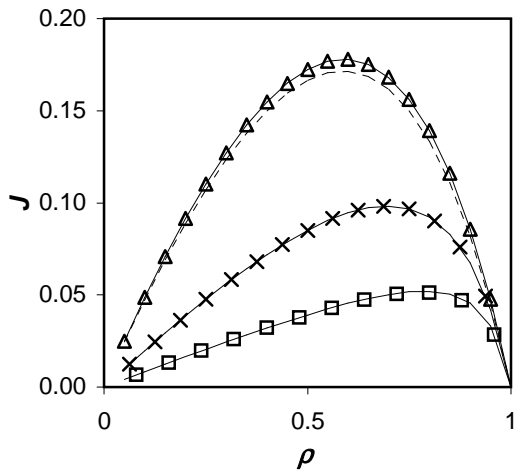}
\caption{Current $J$ vs. coverage density $\rho =M\ell /N$ for
closed systems. Symbols are Monte Carlo results, solid curves are
predicted values from Eqn.\ (\ref{J}), and broken curve is
prediction from Eqn.\ (\ref {MGPJvsrho}). Triangles are for $\ell
=2$, $N=40$, $\times$'s for $\ell =5$, $N=80$, and squares for
$\ell =12$, $N=150$. $J$ was determined by averaging over $ 1.2
\times 10^{5}$ MCS and 100 identical systems after steady state
was reached.} \label{fig:Jvsrho}
\end{figure}

Containing less information, but easier to grasp, is the average
current $ \bar{J}\equiv \sum Jp\left( J\right) $. Its computation
is somewhat easier, since it is $\sum_{H=1}^{M}HZ(H;\tilde{M}
,M)/NZ(\tilde{M}.M)$, with the numerator easily gleaned from
$\left.
\partial _{\eta} W_{M}(\zeta ,\eta )\right| _{\eta =1}.$ The result is
\begin{equation}
\bar{J}=\frac{M}{N}\frac{\tilde{M}}{\tilde{M}+M-1}\,\,. \label{J}
\end{equation}
As we will see, the dependence of this average current on the
density of particles plays a central role. Expressing this
quantity in terms of $\rho _r\equiv M/N$, we have $\bar{J}=\rho
_r\left( 1-\ell \rho _r\right) /\left[ 1-\left( \ell -1\right)
\rho _r-1/N\right] .$ An appealing form, which displays both the
intensive nature of $J$ and its underlying particle-hole symmetry,
is
\begin{equation}
\bar{J}=\frac{\rho _r\rho _h}{\rho _s-1/N}\,\,.  \label{J-sym}
\end{equation}
A third form, frequently referred to in the literature as the
``current-density relationship,'' is writing $\bar{J}$ as a
function of $\rho $, the coverage density ($\rho \in \left[
0,1\right] $):
\begin{equation}
\bar{J}(\rho )=\frac {\rho} {\ell} \frac{1-\rho }{1-\rho +\rho
/\ell -1/N}\,\,. \label{Jvsrho}
\end{equation}
In the limit $N\rightarrow \infty $, this result was first
presented by MacDonald \textit{et al.} \cite{MGP}. A
generalization of the well-known expression for $\ell =1$ (i.e.,
$\bar{J}=\rho (1-\rho )$), this $\bar{J} (\rho )$ is no longer
symmetric about $\rho =1/2$. Instead, the optimal density
increases from $1/2$ to $\sqrt{\ell }\left/ \left( 1+\sqrt{\ell }
\right) \right. $, while the maximum current is lowered from $1/4$
to $\left( 1+\sqrt{\ell }\right) ^{-2}$. As these quantities will
appear frequently, we will denote them by
\begin{equation}
\hat{\rho}\equiv \frac{\sqrt{\ell }}{1+\sqrt{\ell }}\quad
\text{and}\quad \hat{J}\equiv \left( 1+\sqrt{\ell }\right)
^{-2}\,\,.  \label{hats}
\end{equation}
To appreciate this shift graphically, we present, in Figure
\ref{fig:Jvsrho}, both analytic and simulation results. In
addition, to show the effects of the finite size corrections (due
to the $ 1/N$ term), we include a curve of the limiting form
\cite{MGP, LC}
\begin{equation}
\bar{J}\rightarrow \frac{\rho _{r}\rho _{h}}{\rho _{s}}
\label{MGPJvsrho}
\end{equation}
for the $N=40$ case.

In connection with these expressions, a conclusion for \emph{\
conditional }probabilities can be drawn. Since $\bar{J}$ is
precisely the joint probability of finding a ``covered''-hole
pair, we see that $\rho _{r}/\rho _{s}$ is the probability that
site $i$ is covered, given that site $i+1$ is empty and,
similarly, $\rho _{h}/\rho _{s}$ is the probability that site
$i+1$ is empty, given that site $i$ is covered. These conditional
probabilities will play a role in our understanding of the
behavior of open systems.

In addition to the average current, we can also compute its
fluctuations exactly.
\begin{equation*}
\Delta J^{2}\equiv \sum J^{2} p \left( J \right) -\bar{J}^{2}=\frac{\bar{J}^{2}}{\rho
_{s}N} \,\,.
\end{equation*}
Typical of non-critical thermodynamic systems, in which the
(fractional) deviations are $O\left( N^{-1/2}\right) $, the full
distribution $p(J)$ approaches the standard Gaussian form: $\exp
\left[ -\frac{N}{2}\rho _{s}\left( J/\bar{J}-1\right) ^{2}\right]
$. Indeed, this is the form we see in the example shown in Figure
\ref{fig:curDist} above.

\begin{figure}
\includegraphics[clip=true]{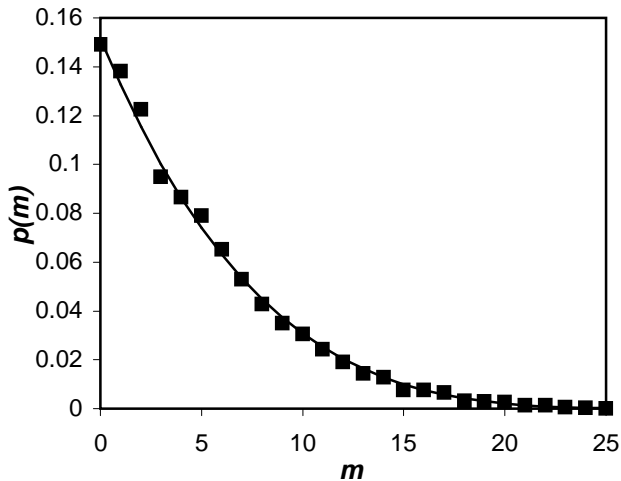}
\caption{\label{fig:headdist} Distribution of head spacings $p(m)$
for $N=100$, $M=6$, and $\ell =12$. Curve is prediction from Eqn.\
(\ref{headspace}), and squares are simulation values. 100 lattices
were simulated to steady state, and head spacing $m$ for any
particles on the first lattice sites was determined every 100 MCS
thereafter for $1.2\times 10^{5}$ MCS.}
\end{figure}

Finally, we turn to another quantity of interest: the statistics
of ``headway.'' Since there are $M$ gaps in each configuration and
there are $Z( \tilde{M},M)$ configurations, we have a total of
$MZ(\tilde{M},M)$ gaps. Out of these, we wish to compute the
number of gaps which contain precisely $m$ holes, which we denote
by $Z(m;\tilde{M},M)$. From its definition
\begin{equation*}
Z(m;\tilde{M},M)\equiv \sum_{\left\{ h_{\alpha }\right\} }\delta
\left( \tilde{M},\sum_{\alpha }h_{\alpha }\right) \left(
\sum_{\alpha }\delta \left( m,h_{\alpha }\right) \right) ,
\end{equation*}
we find the associated generating function
\begin{eqnarray*}
W(\zeta ,\eta )&\equiv& \sum_{\tilde{M},m}Z(m;\tilde{M},M)\zeta
^{\tilde{M}
}\eta ^{m} \\
&=&M\left( \frac{1}{1-\zeta \eta }\right) \left( \frac{1}{1-\zeta
} \right) ^{M-1},
\end{eqnarray*}
leading to the distribution of head spacings
\begin{eqnarray}
p(m;\tilde{M},M)&\equiv& \frac{Z(m;\tilde{M},M)}{MZ(\tilde{M},M)}
\nonumber \\
&=&\frac{ \binom{ \tilde{M}+M-m-2}{\tilde{M}-m}}
{\binom{\tilde{M}+M-1}{\tilde{M}}} \,\,.  \label{headspace}
\end{eqnarray}
In the limit of large $N$, this expression simplifies to
\begin{equation*}
p(m)\rightarrow \frac{\rho _r}{\rho _s}\left( \frac{\rho _h}{\rho
_s}\right) ^m\,\,.
\end{equation*}
This distribution is reproduced faithfully in Monte Carlo
simulations, as shown by an example in Figure \ref{fig:headdist}.
It is easy to understand this result intuitively if we regard
$\rho _h/\rho _s$ as the probability of having a single hole in
the headway. With independent hole statistics, we have
$p(m)\propto \left( \rho _h/\rho _s\right) ^m$. The average number
of holes in the headway and the associated standard deviation can
be computed easily: $\rho _h/\rho _r$ and $\sqrt{\left(\rho_h /
\rho_r \right)^2 + \rho_h / \rho_r }$, respectively.

Though the system considered in this section, particles traveling
on a ring, bears little resemblance to the translation process, it
is sufficiently simple for us to derive a number of exact results.
Apart from their own interest, these results provide crucial
insights, such as the current-density relationship, for
postulating appropriate equations in a coarse grained, mean field
approach to the physical problem at hand.

\section{Extended objects in open systems}
\label{sec:open}

In this section, we attempt the next step towards a realistic
model for protein synthesis by considering systems with open
boundaries.  The section is organized as follows. We describe the
problem and introduce some terminology. Next, we study the effects
of the open boundaries alone, by analyzing the continuum limit of
a \textit{symmetric} exclusion process. Finally, we consider the asymmetric
exclusion process with open boundaries. We present the phase
diagram found from simulations and an extremal principle analysis,
and we show the ability of the continuum limit to predict steady
state density profiles.

In the open system, the first site ($i=1$) is no longer ``in front
of'' the $i=N$ site. Instead, a particle (of extent $\ell $) will
be placed at $i=1$ with probability $ \alpha $ (previously labeled
by $p_{0}$), \emph{provided }all\emph{\ }the first $\ell $ sites
are empty. This models the initiation process. As for elongation,
we continue to restrict ourselves to uniform rates here. In the
language of TASEP, every (randomly) chosen particle \emph{will
move} by one site (increasing $i$ by unity) if it has some
headway. For particles on the last $\ell$ sites, there is no
hinderance, so they will always move if chosen. Finally, to
simulate termination, a particle on the $N^{th}$ site will be
removed from the system with rate $\beta $ (previously labeled by
$ p_{N}$). To repeat, in a Monte Carlo step (MCS), $M+1$ particles
are chosen randomly (in sequence) to attempt a move. They are
selected from a pool including the $M$ particles on the lattice
plus an unbound particle that may initiate.

Since this system is no longer closed, $M$ and $\tilde{M}$ are
fluctuating quantities. Of course, their average values will be
controlled by the rates $ \alpha $ and $\beta $. Our goal is to
find the average densities and the average current of such a
model, as functions of $(\alpha ,\beta )$. In the $ \ell =1$ case,
it is known that the system exhibits three different ``phases'' as
$\alpha $ and $\beta $ are varied, only one of which resembles the
closed system above (in the sense that the current approaches
$\hat{J}$ for large $ N $). Beyond overall averages, we seek the
density profile, which will not only be non-trivial, but which
also displays drastically different properties as we move about in
the $\alpha $-$\beta $ plane. The main goal of this paper is to
study the effects of $\ell >1$ on both the phase diagram and these
density profiles. Unfortunately, $P^{\ast }\left( C\right) $ for a
system with open boundaries is not known in general. Even for the
$\ell =1$ case, only a limited set of quantities may be computed
exactly. To make progress, we resort here to a more
phenomenological approach, in the spirit of hydrodynamics or
Lanudau-Ginzburg free-energy functionals. Considering
coarse-grained densities and the continuum limit, we postulate
equations of motion, based on some of the properties of the closed
system. In principle, such equations can be ``derived'' from the
master equation for $P\left( C,t\right) $, using the mean-field
approximation (i.e., ignoring all correlations). In a sense, this
is also the approach of MacDonald \textit{et al.} \cite{MGP},
except that they focused on $n_{i}^{\left( L\right) }\left(
t\right) $ and $n_{i}^{\left( 0\right) }\left( t\right) ,$ the
probabilities for site $i$ to be occupied by a ribosome and a
hole, respectively, at time $t$. By keeping the spatial
co-ordinate discrete, they faced difference equations and
succeeded in finding solutions only numerically. In contrast, by
using a continuous spatial co-ordinate, we have ordinary
differential equations instead, giving us \emph{analytic}
solutions. However, lost in the notion of ``coarse-graining'' are
the period $\ell $ structures which feature prominently behind
``blockages.'' Nevertheless, our approach appears to capture the
essential (gross) features of these systems.

In our heuristic approach, we imagine $N$ to be large enough to
justify taking the continuum limit, i.e., replacing the discrete
site label $i$ by a continuous co-ordinate: $x$. For simplicity,
define
\begin{equation*}
x\equiv i/N
\end{equation*}
so that $x$ lies within the unit interval. (If physical units of
length are desired, we may introduce $a$ as the lattice spacing,
corresponding to the length of a codon, i.e., three bases of the
mRNA. Then $L_{mRNA}\equiv Na$ would be the length of the mRNA in
question.) Similarly, continuous \emph{local} densities will take
the place of the discrete occupation variables. For example, the
coverage density $\rho \left( x\right) $ will be used instead of
$n_{i}$. Since the maximum occupancy is unity, the hole density is
just
\begin{equation*}
\rho _{h}\left( x\right) =1-\rho \left( x\right) \,\,.
\end{equation*}
Following our considerations above, we also define the ribosome
density by
\begin{equation*}
\rho _{r}\left( x\right) =\rho \left( x\right) /\ell \,\,.
\end{equation*}
Note that, despite the continuum limit, these equations display
the meaning of $\ell $, which serves as a measure of the ``size''
(or ``extent'') of ribosome. Of course, we are also interested in
their time dependence, so that we must consider $\rho \left(
x,t\right) $ in general. Now, ribosomes rarely detach from the
mRNA during elongation, so that we are justified in regarding
these densities as \emph{conserved} fields. Thus, the appropriate
equation of motion is the continuity equation, i.e.,
\begin{equation}
\frac{\partial }{\partial t}\rho _{r}\left( x,t\right)
=-\vec{\nabla}\cdot \vec{J}_{r}=-\frac{\partial }{\partial
x}J_{r}\left( x,t\right) \,\,, \label{ContEq}
\end{equation}
where $J_{r}\left( x,t\right) $ is the local (ribosome) current.
Our first task is to find how this current depends on the (local)
density $\rho
_{r}\left( x,t\right) $, i.e., to find the functional form for $J_{r}\left[ \rho _{r}
\right] $. Then we will arrive at an acceptable equation for $\rho
_{r}$. To find the steady state profile $\rho _{r}^{\ast }\left(
x\right) $, which satisfies $\partial _{t}\rho _{r}^{\ast }=0$, we
see that this state is associated with a constant
($x,t$-independent) current $J_{r}^{\ast }$. Therefore, our
problem consists of finding the solution to
\begin{equation*}
J_{r}\left[ \rho _{r}^{\ast }\right] =constant\,\,,
\end{equation*}
subject to the appropriate boundary conditions. To keep the
notation simple, from here on we will drop the subscript $r$ and
write $J$ or $ J\left( x,t\right) $ for the local current (as well
as $J^{*}$ for the constant current in the stationary state).

\subsection{The case of free diffusion and an effective density variable for
extended objects}

To show how we build an appropriate equation for the TASEP, let us
begin with a study of the effects of open boundaries \emph{alone}.
In other words, let us consider a \emph{symmetric} exclusion
process for extended objects, i.e., a system of large particles
diffusing freely on a line, subjected to the excluded volume
constraint and the controls $\alpha ,\beta $. For simulations, a
randomly chosen particle is moved one site forward or backward
with equal probability ($0.5$), provided it does not run into its
neighbor. The only exception is the first particle, which is
prohibited from jumping backwards into the ``source.''

\begin{figure}[tbp]
\includegraphics[clip=true]{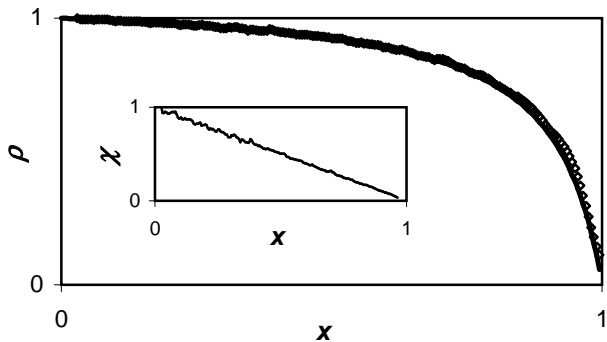}
\caption{Density profile for free diffusion in a system with
$N=200$ and $\ell=12$ from $\rho=1$ to $\rho=0$. Symbols are Monte
Carlo data. $100$ identical systems were simulated in parallel for
$10^{5}$ MCS to reach steady state, and then density profiles were
collected every 100 MCS for an additional $1.2\times 10^{4}$ MCS. Curve is
predicted density profile. Inset shows the effective particle
density profile $\chi$, constructed from the simulation data. The
current observed in simulations was $J^{\ast }=0.00252$, which is
close with the expected value of $0.0025=D\left[ \chi \left(
0\right) -\chi \left( 1\right) \right]$ for $D=1/2N$.}
\label{fig:freediff}
\end{figure}

For the $\ell =1$ case ($\rho =\rho _{r}$), the steady state
profile is trivially linear:
\begin{equation*}
\rho ^{\ast }\left( x\right) =\left( 1-x\right) \rho ^{\ast
}\left( 0\right) +\left( x\right) \rho ^{\ast }\left( 1\right)
\end{equation*}
with current
\begin{equation*}
J^{\ast }=\frac{1}{N}\left[ \rho ^{\ast }\left( 0\right) -\rho
^{\ast }\left( 1\right) \right] \,\,.
\end{equation*}
(Note that, since the current is controlled by the gradient of the
local density only, it vanishes necessarily in the $N\rightarrow
\infty $ limit. So, we must keep the $N$ explicitly here.)
Meanwhile, the boundary densities are fixed by matching the
injection/depletion rates to the internal current:
\begin{equation*}
\alpha \rho _{h}^{\ast }\left( 0\right) =\alpha \left[ 1-\rho
^{\ast }\left( 0\right) \right] =J^{\ast }=\beta \rho ^{\ast
}\left( 1\right) \,\,,
\end{equation*}
and our problem is completely solved. These well known results can
be traced to the fact that Eqn.\ (\ref{ContEq}) assumes the form
of the simple diffusion equation $\partial _{t}\rho \propto
\partial _{x}^{2}\rho $, since $J\left[ \rho \right] \propto
-\nabla \rho $. Though there are many ways to arrive at this
result, it is less clear how to generalize it to the case of
extended objects. In particular, as displayed in Figure
\ref{fig:freediff}, the profile from simulations (for $N=200,\ell
=12,\alpha =\beta =1$) is far from linear! Here, we will show that
there is a natural generalization for the (particle) density in
the case of extended objects and that its profile is again linear.

We proceed by returning to the basics, starting with the current
$J_{r}$ being proportional to both the conductivity (mobility) and
the drive. For the former, we take the result from the previous
section. There, the drive is constant, so that the right hand side
of Eqn.\ (\ref{MGPJvsrho}), $\rho_r\rho_h/\rho_s$, can be
interpreted as the conductivity or mobility (neglecting the
finite-size effect term from the closed system). Meanwhile, the
drive for free diffusion should be the gradient of a pressure
$\mathcal{P}$, so that
\begin{equation}
J_{r}\left[ \rho \right] =D\left[ \frac{\rho _{r}\rho _{h}}{\rho
_{s}}\right] \left[ -\nabla \mathcal{P}\right] \,\,,
\label{freediffcurr}
\end{equation}
where $D$ is a constant, to be fitted to data. Since we have
scaled the system size to unity, we should keep in mind that $D $
is a quantity of $O\left( 1/N\right) $, to be consistent with the
continuum limit. For the pressure $\mathcal{P}$, we follow
the standard route of statistical mechanics and write
\begin{equation*}
\mathcal{P}=\frac{\delta \mathcal{H}}{\delta \rho _{r}} \,\,,
\end{equation*}
where $\mathcal{H}$ is a free energy functional (e.g., the
Landau-Ginzburg ``Hamiltonian'' in case of the ordinary lattice
gas). Here, we have a non-interacting system, so that a reasonable
form for $\mathcal{H}$ is just the entropy \cite{entropynote}:
\begin{equation*}
\mathcal{H}=\int dx\left[ \rho _{r}\ln \rho _{r}+\rho _{h}\ln \rho
_{h}-\rho _{s}\ln \rho _{s}\right] \,\,.
\end{equation*}
Using
\begin{equation*}
\rho _{h}=1-\ell \rho _{r}\quad ,\quad \rho _{s}=1-\left( \ell
-1\right) \rho _{r}
\end{equation*}
and carrying out the steps, we arrive at a \emph{modified}
diffusion equation:
\begin{equation}
\partial _{t}\rho _{r}=D\partial _{x}\left[ \rho _{s}^{-2}\partial _{x}\rho
_{r}\right] \,\,. \label{diff}
\end{equation}
Note that, for $\ell >1$, the effective ``diffusion constant,''
$D/\rho {}_{s}^{2}$, is \emph{density-dependent}. Thus, the steady state
profile will not be linear in $x$. Instead, it satisfies
\begin{equation}
\frac{D}{\left( \rho _{s}^{\ast }\right) ^{2}}\frac{\partial \rho
_{r}^{\ast }}{\partial x}=-J^{\ast }\,\,,  \label{SS1}
\end{equation}
which will definitely lead to non-linear profiles.

Although we have derived this diffusion equation via
phenomenological techniques, we speculate that it could be
obtained through a more rigorous derivation.  Specifically, if a
matrix product description of the $\ell>1$ system becomes
available, the methods of Derrida \textit{et al.} \cite{DLS1} may
extend to calculate a large deviation functional, which would play
the role of a nonequilibrium free energy for this system. Should
such a functional exist, we expect that our diffusion equation
would result from a variation of this functional.

To find $\rho _{r}^{\ast }$ explicitly, notice that, due to $\rho
_{s}=1-\left( \ell -1\right) \rho _{r}$, the left hand side is
just a simple derivative of $1/\rho _{s}^{\ast }$. Thus, the
profile $1/\rho _{s}^{\ast }\left( x\right) $ will be
\emph{linear.} However, $\rho _{s}$ does not reduce to a sensible
variable as $\ell \rightarrow 1$, leading us to define
\begin{equation*}
\chi \equiv \frac{1}{\ell -1}\left[ \frac{1}{\rho _{s}^{\ast
}}-1\right] = \frac{\rho ^{\ast }}{\ell -\left( \ell -1\right)
\rho ^{\ast }}\,\,.
\end{equation*}
Not only does $\chi $ lie in the unit interval; it reduces to the
usual particle density $\rho ^{\ast }$ in the limit $\ell
\rightarrow 1$. Most importantly, it is uniquely related to $\rho
^{\ast }$ for all $\ell $. First introduced in \cite{FA}, $\chi $
is the ``natural'' generalization of the particle density for
extended objects. We will refer to it as the effective particle
density (EPD). Meanwhile, a hole remains a single-site entity, so
its density needs no modification. In terms of the EPD, the
various densities are
\begin{eqnarray*}
\rho _{s}^{\ast }=\frac{1}{1+\left( \ell -1\right) \chi };\,\,
\nonumber \\
\rho_{r}^{\ast }=\frac{\chi }{1+\left( \ell -1\right) \chi };\,\, \nonumber \\
\rho_{h}^{\ast }=\frac{1-\chi }{1+\left( \ell -1\right) \chi
}\,\,\,\,
\end{eqnarray*}
so that $\chi $ is just
\begin{equation}
\chi\equiv\rho _{r}^{\ast }/\rho _{s}^{\ast }. \label{chidef}
\end{equation}
The current-density relationship (Eqn.\ \ref{MGPJvsrho}) for the
steady state in the \emph{ring} is again a simple product:
\begin{equation}
\bar{J}_{ring}=\chi \rho _{h}\,\,.  \label{Jbar}
\end{equation}
Note that this product will serve as the \emph{mobility} in our
mean field approach here and can be expressed in terms of $\chi $
alone:
\begin{equation*}
\chi \rho _{h}=\frac{\chi \left( 1-\chi \right) }{1+\left( \ell
-1\right) \chi }\,\,.
\end{equation*}
The main advantage of using the EPD is the re-emergence of the
familiar combination $\chi \left( 1-\chi \right) $ in the
numerator. Finally, corresponding to the optimal density (Eqn.\
\ref{hats}), we have
\begin{equation*}
\hat{\chi}=\frac{1}{1+\sqrt{\ell }}\,\,,
\end{equation*}
a quantity which plays a significant role in systems with open
boundaries. Note that the corresponding hole density
($1-\hat{\rho}$) also assumes the same value, so that $\hat{J}$ is
just $\hat{\chi}^{2}$. The underlying ``particle-hole'' symmetry
generalizes to one under exchange of $\chi \Leftrightarrow $ $\rho
_{h}$, or
\begin{equation*}
\chi \Leftrightarrow \frac{\left( 1-\chi \right) }{1+\left( \ell
-1\right) \chi }\,\,,
\end{equation*}
or, in terms of the more physical $\rho $,
\begin{equation*}
\rho  \Leftrightarrow \frac{1-\rho }{1-\rho +\rho /\ell }\,\,.
\end{equation*}

We now find that Eqn.\ (\ref{SS1}) simplifies to the familiar
equation from ordinary diffusion:
\begin{equation*}
D\frac{\partial \chi }{\partial x}=-J^{\ast }\,.
\end{equation*}
As a result, the profile in terms of $\chi $ is again linear in
$x$ \cite{linearnote}. For completeness, we write the solution:
\begin{equation*}
\chi \left( x\right) =\left( 1-x\right) \chi \left( 0\right)
+\left( x\right) \chi \left( 1\right) \,\,,
\end{equation*}
with
\begin{equation*}
\chi \left( 0\right) -\chi \left( 1\right) =J^{\ast }/D\,\,.
\end{equation*}
The explicit solution ensues once the constraints of
injection/depletion rates are imposed. Note that, in general, we
would have $J^{\ast }\propto D$, so that the current inherits the
$O\left( 1/N\right) $ from $D$, in contrast to the $O\left(
1\right) $ behavior for systems with non-zero drive.

To close, we illustrate how well the theory agrees with data by
showing the analytic, non-linear profile of $\rho ^{\ast }\left(
x\right) $ with boundary conditions $\rho \left( 0\right) =1$,
$\rho \left( 1\right) =0$:
\begin{equation*}
\rho ^{\ast }\left( x\right) =\frac{1-x}{1-\left( 1-1/\ell \right)
x}\,\,,
\end{equation*}
as a curve in Figure \ref{fig:freediff}.

\subsection{TASEP in open systems}

We now turn to the other extreme case, where backward jumps are
completely excluded (TASEP), generalizing the model of \cite{DDM}
to extended objects. With open boundaries, this model incorporates
another essential feature of translation. Our interest is again
the average current and density profile, as a function of $\alpha $ and
$\beta $, the rates of initiation and termination.

\subsubsection{Phase diagram from extremal principle and simulations}

Although we are unable to generalize the methods of \cite{DDM, SD}
to arrive at exact solutions for an open system with particles
covering $\ell >1$ sites, we find that the phase diagram
determined by the mean field techniques of \cite{MG, LC} can be
understood also by extending the extremal principle, an hypothesis
first proposed by Popkov and Sch\"{u}tz \cite{PS} and later
exploited successfully by others (e.g., \cite{Hager1, Hager2}). In
this approach, the open boundaries are regarded as connections to
reservoirs with appropriate densities, so that by keeping the same
jump rates as in the bulk, $\alpha $ and $\beta $ are realized.
Defining $\rho _{-}$ and $\rho _{+}$ as the reservoir densities at
the initiation and termination boundaries, respectively, the
extremal principle relates the current in the open system to the
$J[\rho ]$ for a closed, periodic system with the same bulk
dynamics \cite{inflimitnote}:
\begin{equation*}
J=\left\{
\begin{tabular}{ll}
$\max \,J[\rho ]$ & $\text{for\ }\rho _{+}<\rho <\rho _{-}$ \\
$\min \,J[\rho ]\text{ }$ & $\text{for }\rho _{-}<\rho <\rho _{+}$
\end{tabular}
\right. \,\,.
\end{equation*}
Unfortunately, there is no prescription for finding $\rho
_{-},\rho _{+}$ from $\alpha ,\beta $ in general. Exploiting
results from the exactly
soluable $\ell =1$ case (where $\rho _{-}=\alpha $ and $\rho _{+}=1-\beta $
) and from previous studies of the $\ell >1$ case \cite{MG, LC}, we argue in favor of
\begin{equation}
\rho _{-}(\alpha )=\frac{\ell \alpha }{1+\alpha \left( \ell
-1\right) }\quad \text{and}\quad \rho _{+}(\beta )=1-\beta .
\label{reservoirs}
\end{equation}
These reservoir densities can be understood as follows. Recall
that, for a closed system, the probability for site $i-1$ to be
filled, \emph{given} that site $i$ is empty, is $ \rho _{r}/\rho
_{s}=\left( \rho /\ell \right) /\left( 1-\rho +\rho /\ell \right)
$. We now argue that when the first site of an open system is
empty and will be filled with probability $\alpha $, it can be
thought of as being coupled to a reservoir of the appropriate
density, i.e., $\rho _{-}$ such that $ \alpha =\left( \rho
_{-}/\ell \right) /\left( 1-\rho _{-}+\rho _{-}/\ell \right) $.
Solving for $\rho _{-}$ leads to the expression above for $\rho
_{-}(\alpha )$. The expression for $\rho_{+} \left( \beta \right)$
is not readily explained by similar arguments, so discussion of
its origin will be deferred until the following section. An
important feature associated with this choice is that the current
is symmetric under $\alpha \Leftrightarrow \beta $, as observed in
simulations. Most importantly, these choices lead to phase
diagrams in good agreement with Monte Carlo data.

\begin{figure}[tbp]
\centering
\includegraphics[clip=true]{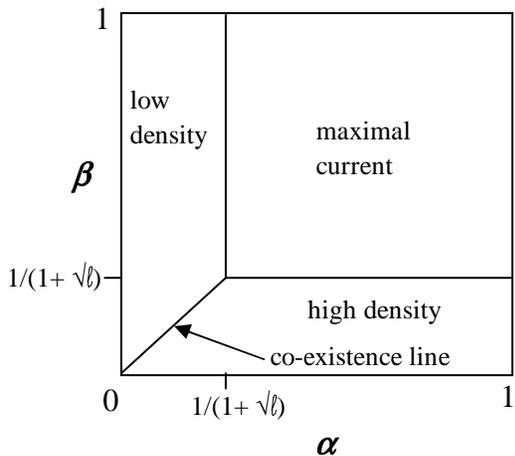}
\caption{Phase diagram predicted by the extremal principle.}
\label{fig:pd}
\end{figure}

\begin{figure}[tbp]
\centering
\includegraphics[clip=true]{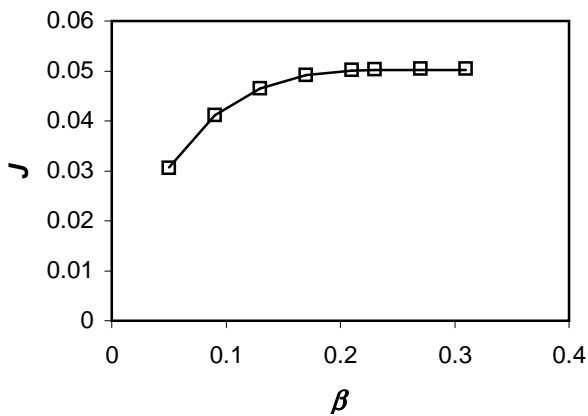}
\caption{Dependence of current on $\beta $ for $\alpha =1$, $ \ell
=12$, and $N=1000$. Symbols are simulation results (determined
from 100 systems simulated in parallel for $1.2 \times 10^{4}$ MCS
after steady state was reached) and curve is the prediction from
Eqn.\ (\ref{Jab}).}\label{fig:Jofb}
\end{figure}

\begin{figure}[tbp]
\centering
\includegraphics[clip=true]{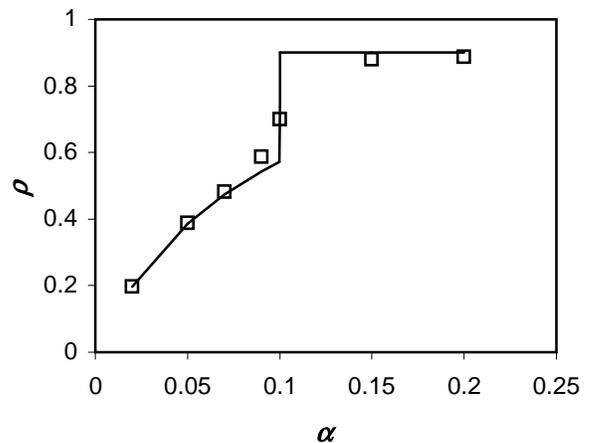}
\caption{Dependence of average coverage density $\rho$ on $\alpha
$ for $\beta =0.1$, $\ell =12$, and $N=1000$. Symbols are
simulation results (determined from 100 systems simulated in
parallel and sampled every 100 MCS for $1.2 \times 10^{4}$ MCS
after steady state was reached) and curve is the prediction from
$\rho _{-}(\alpha )$ and $\rho _{+}(\beta )$ in Eqn.\
(\ref{rho-ab}).} \label{fig:rhoofa}
\end{figure}

By combining these choices with the extremal principle, we find
that, although the result is qualitatively similar to the $\ell
=1$ system, there are quantitative changes for the $\ell >1$ case.
First, there is a shift in the location of transition lines, from
$1/2$ to $\hat{\chi}= 1/\left( 1+\sqrt{\ell }\right) $ (as shown
in Figure \ref{fig:pd}). Then, the current and bulk densities
are also modified:
\begin{equation}
\bar{J}\left( \alpha ,\beta \right) =\left\{
\begin{tabular}{ll}
$\frac{\alpha (1-\alpha )}{1+\alpha (\ell -1)}$ & $\text{for
}\alpha <\beta
<\hat{\chi}\text{ (low density)}$ \\
$\frac{\beta (1-\beta )}{1+\beta (\ell -1)}\text{ }$ & $\text{for
}\beta
\leq \alpha <\hat{\chi}\text{ (high density)}$ \\
$\hat{J}\text{ \ }$ & $\text{for }\alpha ,\beta \geq
\hat{\chi}\text{ (max current)}$\,,
\end{tabular}
\right. \label{Jab}
\end{equation}
and
\begin{equation}
\bar{\rho}\left( \alpha ,\beta \right) =\left\{
\begin{tabular}{ll}
$\rho _{-}$ & $\text{for }\alpha <\beta <\hat{\chi}\text{ (low
density)}$
\\
$\rho _{+}$ & $\text{for }\beta <\alpha <\hat{\chi}\text{ (high
density)}$
\\
$\hat{\rho}\text{ \ }$ & $\text{for }\alpha ,\beta \geq
\hat{\chi}\text{ (max current)}$ \,.
\end{tabular}
\right. \label{rho-ab}
\end{equation}
As examples of how well these predictions fit simulation data, we
show plots of $\bar{J}\left( 1,\beta \right) $ (Figure
\ref{fig:Jofb}) and $\bar{\rho}\left( \alpha ,0.1\right) $ (Figure
\ref{fig:rhoofa}). As expected, when $\alpha $ was rate-limiting
(low density phase), a bulk coverage density of $\rho _{-}(\alpha
)$ was induced, and when $\beta $ was rate-limiting, a bulk
coverage density of $\rho _{+}(\beta )$ was induced. Given the
success of the extremal principle analysis, it would be
appropriate to search for more fundamental theories from which the
extremal principle could be derived.

\begin{figure}[tbp]
\centering
\includegraphics[clip=true]{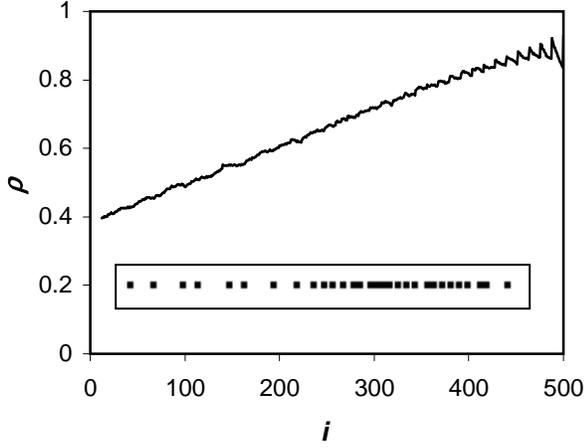}
\caption{Steady state density profile for $\alpha=0.05$,
$\beta=0.05$, $N=500$, $\ell=12$. Profile was obtained from
simulations of 100 systems run to steady state and then sampled
every 100 MCS for $1.2 \times 10^{4}$ MCS. (Nonlinearities near the
termination boundary result from ribosomes tending to ``pile up'' at
$i=N,N-\ell , N-2\ell$, etc. due to $\beta <1$ but with decreasing
correlations as $i$ decreases.)
Inset shows a typical configuration
of this system, with the shock front near the center. (Each
ribosome is represented by a dot.)} \label{fig:1stprofile}
\end{figure}

\begin{figure}[tbp]
\centering
\includegraphics[clip=true]{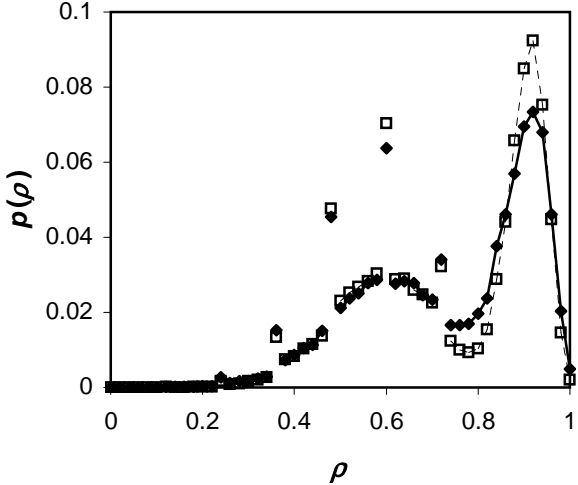}
\caption{Probability distribution of densities in the central 10\%
of a system with $\alpha=0.1$, $\beta=0.1$, $N=1000$, and
$\ell=12$. Actual probability distribution (solid diamonds with
solid curve) was obtained by simulating 100 systems to steady
state, sampling the central densities every 100 MCS for $5 \times
10^{4}$ MCS, and compiling a frequency histogram. Peaks correspond
to the densities $\rho_{-}\left(\alpha\right)$ and
$\rho_{+}\left(\beta\right)$. For comparison, a composite density
distribution (open squares with dashed curve) was assembled from
similar frequency histograms for two closed systems with average
densities $\rho_{-}\left(0.1\right)$ and
$\rho_{+}\left(0.1\right)$. (Again, $N=1000$, $\ell=12$, and 10\%
of the system was sampled.) The composite distribution required a
linear combination of 53\% low density and 47\% high density to
minimize the absolute difference from the actual probability
distribution. See text for details.} \label{fig:1sthist}
\end{figure}

The transition between low and high density phases is clearly
first order, displayed as a jump in the bulk density plot (Figure
\ref{fig:rhoofa}). As in the $\ell =1$ case \cite{KSKS}, domain
wall theory can be used to understand our results. In particular,
for $\alpha =\beta <\hat{\chi}$, we have observed a shock front
(between the low and high density regions) diffusing freely along
the lattice. As a result, the \emph{average} profile is linear, as
shown in Figure \ref{fig:1stprofile}. To appreciate the shock, we
display in the inset a typical configuration (showing a shock)
where each ribosome is represented by a dot. In addition, we
exhibit another aspect of this co-existence by sampling the
average (coverage) density in the central 10\% of a large
($N=1000$) lattice. Compiling a histogram (closed diamonds in
Figure \ref{fig:1sthist}), we find a bimodal distribution typical
of systems at a first order transition. Peaks are expected to
correspond to the densities $\rho_{-}\left(\alpha\right)$ and
$\rho_{+}\left(\beta\right)$. To connect with local fluctuations
in the ``pure'' systems, we compile similar histograms for a
closed system (i.e., 10\% of a ring with $1000$ sites) with
overall density set at $\rho _{-}(\alpha )$ and $\rho _{+}(\beta
)$. For comparison, we show a simple linear combination of such
distributions in Figure \ref{fig:1sthist} (open squares). We
expect correspondence between densities in the window in the open
system, which alternate between approximately
$\rho_{-}\left(\alpha\right)$ and $\rho_{+}\left(\beta\right)$,
and an appropriately weighted average of window densities in the
two closed systems. The deviations can be understood as the
contribution of configurations with the shock in the sampling
window. Roughly, this may occur about 10\% of the time, which is
also the order of magnitude of the deviation from the simple
average of pure systems.

While most of the features presented here were known to MacDonald
\textit{et al.} \cite{MGP, MG} and have also been obtained by Lakatos
and Chou \cite{LC}, our efforts are to go beyond mean field approaches,
showing the different perspective offered by domain wall theory and
the extremal principle. Shock effects on the co-existence line are
particularly well described by this approach.

\subsubsection{Steady state profiles from the continuum approach}

Following an understanding of the phase diagram, we turn to more
details of the system, namely, steady state density profiles. Our
continuum approach again leads us to differential equations. As
considerable insight can be gained by studying simple differential
equations, we believe it is worthwhile to devote a section to the
continuum approach.

Starting with the case of free diffusion, we extend Eqn.\
(\ref{diff}) to the driven system by simply adding an ``Ohmic''
term to the current: $J=\sigma E$, where the drive is associated
with strength $E$ and the conductivity will be the density
dependent factor found before (first used in Eqn.\
\ref{freediffcurr}). Thus, our starting equation is
\begin{equation}
\frac{\partial \rho _{r}}{\partial t}=-\frac{\partial }{\partial
x}J\left( x\right) =\frac{\partial }{\partial x}\left[
\frac{D}{\rho _{s}^{2}}\frac{
\partial \rho _{r}}{\partial x}-\frac{\rho _{r}\rho _{h}}{\rho _{s}}E\right]
\,\,.  \label{drivendiff}
\end{equation}
As before, we suggest that this driven diffusion equation might be
obtained from a variation of an appropriate large deviation
functional, the $\ell=1$ version of which has been derived
previously \cite{DLS2, DLS3}.

As a reminder, our $x$ is actually the \emph{fractional} length
along the mRNA, so that it is dimensionless. Similarly, the
densities are also unitless (e.g., $\rho _{h}\in \left[ 0,1\right]
$), so that $D/E$ carries the information of the real length of
the chain. An estimate of this ratio can be obtained by taking the
na\"{i}ve continuum limit of a discrete hopping model, with the
result $1/2N$. In the Appendix, we will show how this arises from
taking such a limit of the MacDonald \textit{et al.} \cite{MGP}
current equation. For here, we may regard $D/E$ as a
phenomenological parameter. Further, since we are focusing only on
\emph{totally} asymmetric processes, we could just as well set $E$
to unity (as in Eqn.\ \ref{MGPJvsrho}), and measure $J$ and $D$
``in units of $E$.''

Taking these considerations into account, we seek the steady state
profile $ \rho _r^{*}\left( x\right) $, as in Eqn.\ (\ref{SS1}),
by setting the square bracketed terms in Eqn.\ (\ref{drivendiff})
to the steady state current:
\begin{equation*}
D\left( \rho _s^{*}\right) ^{-2}\frac{\partial \rho
_r^{*}}{\partial x}-\rho _r^{*}\rho _h^{*}\left( \rho
_s^{*}\right) ^{-1}=-J^{*}\,\,.
\end{equation*}
In terms of our ``natural'' variable $\chi $ (Eqn.\ \ref{chidef}),
this equation reduces to
\begin{eqnarray}
\chi ^{\prime } &\equiv & \frac{\partial \chi }{\partial
x}
\nonumber \\
&=& \frac{-1}{D\left[ 1+\left( \ell -1\right) \chi \right] }
\left[ \tilde{J}\chi +J^{\ast }-\chi \left( 1-\chi \right) \right]
, \label{SSPequ}
\end{eqnarray}
where $\tilde{J}\equiv J^{\ast }\left( \ell -1\right) $. Thus, we
see that
this $\ell >1$ generalization is quite similar to the $\ell =1$ equation: $
\rho ^{\prime }=\left[ \rho \left( 1-\rho \right) -J^{\ast
}\right] /D$.

Now, the zeros of $\chi ^{\prime }$ will play an important role,
occurring at
\begin{equation}
\chi _{>}\equiv \frac{1-\tilde{J}+R}2\text{ and }\chi _{<}\equiv
\frac{1- \tilde{J}-R}2, \label{roots}
\end{equation}
where
\[
R\equiv \sqrt{\left( 1-\tilde{J}\right) ^{2}-4J^{\ast }}\,.
\]
It is not surprising that the maximal current $\hat{J}$ (i.e.,
$\left( 1+ \sqrt{\ell }\right) ^{-2}$ in our system of units) is a
key player, so that $ R$ can be written in the following form:
\[
R=\left( \ell -1\right) \sqrt{\left( J^{\ast }-\hat{J}\right)
\left( J^{\ast }-\check{J}\right) }\,.
\]
Here, $\check{J}$ $\equiv \left( 1-\sqrt{\ell }\right) ^{-2}$
appears as a natural counterpart to $\hat{J}$. Another advantage
of this form is its relationship to the extremal current
principle. As we will see, for finite systems, $J^{\ast }$ will be
slightly ($O\left( 1/N^{2}\right) $) larger than $ \hat{J}$ for
the maximal current phase, giving us complex roots and profiles
with inflections. On the other hand, for the other two phases,
$J^{\ast }< \hat{J}$, leading to fixed points in the $N\rightarrow
\infty $ limit.

Eqn.\ (\ref{SSPequ}) can be integrated to find an implicit
function for the density profile in each phase. For explicit
solutions, we must specify the boundary conditions:
\begin{eqnarray}
J^{\ast } &=&\alpha \left[ 1-\rho \left( 0\right) \right]   \nonumber \\
J^{\ast } &=&\frac{\beta }{1+\beta \left( \ell -1\right) }\rho
\left( N\right)   \label{JBCs}
\end{eqnarray}
so that
\begin{equation}
\chi \left( 0\right) =\frac{\rho \left( 0\right) /\ell }{1-\rho
\left( 0\right) +\rho \left( 0\right) /\ell }=\frac{\alpha
-J^{\ast}}{\alpha +\left( \ell -1\right) J^{\ast}}  \label{chiBC0}
\end{equation}
and
\begin{eqnarray}
\chi \left( 1\right) &=&\frac{\rho \left( N\right) /\ell }{1-\rho
\left(
N\right) +\rho \left( N\right) /\ell } \nonumber \\
&=&\frac{\left[ 1+\beta \left( \ell -1\right) \right]
J^{\ast}}{\ell \beta -\left( \ell -1\right) \left[ 1+\beta \left(
\ell -1\right) \right] J^{\ast}}\;. \label{chiBC1}
\end{eqnarray}

The boundary conditions can be understood as follows. From $\rho
_{-}\left( \alpha \right) $ in Eqn.\ (\ref{reservoirs}), we can
show that the effective reservoir at the initiation boundary has
an EPD of $\chi _{-}\left( \alpha \right) =\alpha $. Eqn.\
(\ref{Jbar}) then implies the initiation boundary condition in
Eqn.\ (\ref{JBCs}). Particles on the final $\ell $ lattice sites
experience no steric hindrance, so the current through the final
$\ell $ sites is just
\begin{eqnarray*}
J^{\ast } &=&\rho _{r}\left( i\right) \text{ \ \ \ \ for }i=N-\ell
+1,\ldots
,N-1 \\
J^{\ast } &=&\beta \rho _{r}\left( N\right) \;.
\end{eqnarray*}
These relations can be used to express $\rho \left( N\right)
=\sum_{i=N-\ell +1}^{N}\rho _{r}\left( i\right) $ in terms of
$J^{\ast }$, leading to the termination boundary condition in
Eqn.\ (\ref{JBCs}). The boundary conditions for $\chi $ will serve
to fix the constant of integration as well as $J^{\ast }$, which
is still an unknown in Eqn.\ (\ref{SSPequ}). (The current thus
determined is expected to match closely the current predicted by
Eqn.\ \ref{Jab}.) We discuss the various phases separately.

For smaller values of $J^{\ast }$ (i.e., $J^{\ast }<\hat{J}$),
$\chi \left( x\right) $ has fixed points at $\chi _{>}$ and
$\chi_{<}$. Boundary conditions determine whether the steady state
profile approaches $\chi _{<}$ (low density) or $\chi _{>}$(high
density). Each solution corresponds to part of the phase diagram
shown in Figure \ref {fig:pd}. There is a ``kink'' solution when
$\chi \left( 0\right) ,\chi \left( 1\right) \in \left( \chi
_{<},\chi _{>}\right) $ with $\chi \left( 0\right) \gtrapprox \chi
_{<}$ and $\chi \left( 1\right) \lessapprox \chi _{>}$ that
corresponds to the first order transition line.

In the high density phase ($\beta <\alpha $) where $\chi >\chi
_{>}$, the density profile is given implicitly by
\begin{widetext}
\begin{equation*}
\left[ 1+\left( \ell -1\right) \chi _{>}\right] \ln \left(
\frac{\chi -\chi _{>}}{\chi _{0}-\chi _{>}}\right) -\left[
1+\left( \ell -1\right) \chi _{<} \right] \ln \left( \frac{\chi
-\chi _{<}}{\chi _{0}-\chi _{<}}\right) =- \frac{Rx}{D}
\end{equation*}
\end{widetext}
(where $\chi _{0}\equiv \chi \left( 0\right) $ is the EPD at
initiation). It may be noted that as $N \to \infty$, the bulk
density and termination boundary density both approach $\chi_{>}$.
Setting $\chi \left( 1 \right) = \chi_{>}$ and eliminating
$J^{\ast}$ from the termination boundary condition (Eqn.\
\ref{chiBC1}) and the definition of $\chi_{>}$ (Eqn.\ \ref{roots})
yields a bulk $\chi$ of $\left( 1-\beta \right) / \left[ 1+\beta
\left( \ell -1 \right) \right]$. We expect the bulk density in the
high density phase to match the termination reservoir density.
Indeed, the bulk $\chi$ value calculated here is consistent with
the reservoir density $\rho_{+} =1-\beta$ given above (Eqn.\
\ref{reservoirs}).

Though it is impossible to write a closed form for $\chi \left(
x\right) $ in general, a convenient expression is
\begin{equation}
\chi =\chi _{>}+\left( \chi _{0}-\chi _{>}\right) \exp \left[ -\mu
_{+}x \right] \left( \frac{\chi -\chi _{<}}{\chi _{0}-\chi
_{<}}\right) ^{\gamma } \label{chiHD}
\end{equation}
where
\[
\mu _{+}\equiv \frac{R}{D\left[ 1+\left( \ell -1\right) \chi
_{>}\right] }
\]
and
\[
\gamma =\frac{1+\left( \ell -1\right) \chi _{<}}{1+\left( \ell
-1\right) \chi _{>}}\,.
\]
Note that $R$, $\chi _{<}$, and $\chi _{>}$ can be conveniently
estimated using the $J^{\ast}$ value given by Eqn.\ (\ref{Jab}).
The advantage here is that the unknown factor $\left( \frac{\chi
-\chi _{<}}{\chi _0-\chi _{<}} \right) ^\gamma $ displays only
limited variation, since $\chi $ lies outside the interval $\left[
\chi _{<},\chi _{>}\right] $. Thus, we expect to find better and
better approximations by exploiting an \emph{iterative} scheme.
Starting with the first iteration
\[
\chi ^{(1)}=\chi _{>}+\left( \chi _0-\chi _{>}\right) \exp \left[
-\mu _{+}x\right] \,\,,
\]
this procedure involves substituting repeatedly for the $\chi $ in
the unknown factor of Eqn.\ (\ref{chiHD}), namely,
\[
\chi ^{(k)}=\chi _{>}+\left( \chi _0-\chi _{>}\right) \exp \left[
-\mu _{+}x\right] \left( \frac{\chi ^{(k-1)}-\chi _{<}}{\chi
_0-\chi _{<}}\right) ^\gamma \,\,.
\]

\begin{figure*}[tbp]
\centering
\includegraphics[clip=true]{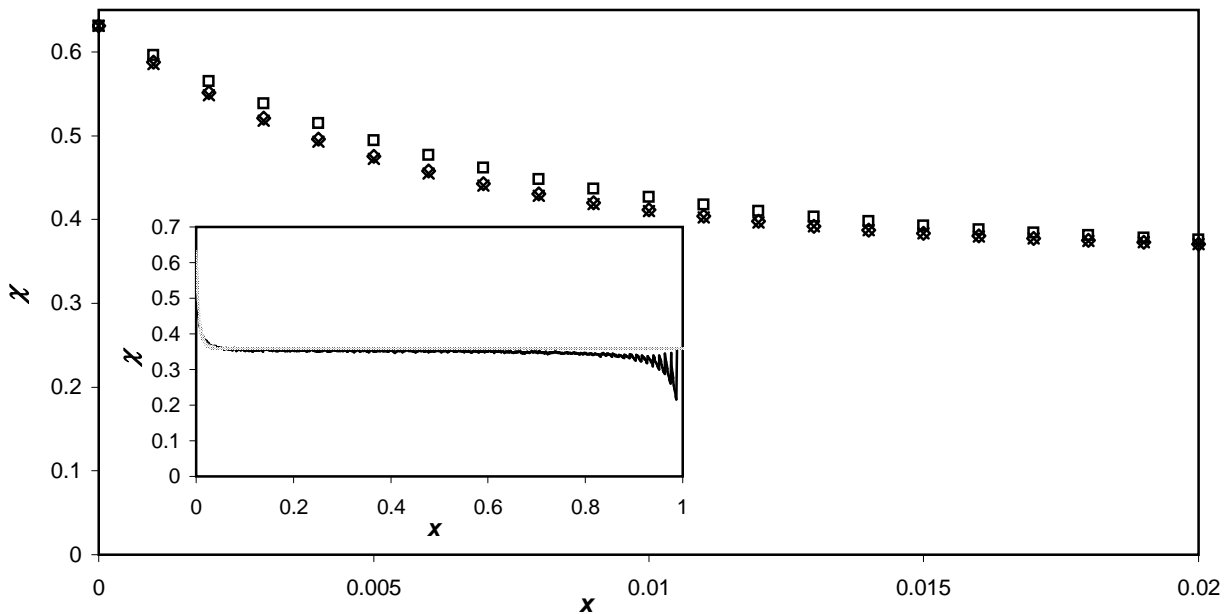}
\caption{Rapid convergence of iterative method to
$\chi\left(x\right)$ for small $x$. (Larger $x$ values are omitted
because all iterations of $\chi$ are nearly identical.)
$\chi^{(1)}$, $\chi^{(2)}$, and $\chi^{(3)}$ are represented by
squares, triangles, and $\times$'s, respectively. The system
considered here is $\alpha=1$, $\beta=0.13$, $N=1000$, $\ell=12$.
$D$ was set to $1/3N$ to obtain a good fit. Inset compares the
actual steady state $\chi$ from simulations (bold curve) with the
predicted $\chi^{(3)}$ (lighter curve). In the profile from
simulations, particle depletion near the termination end can be
seen.} \label{fig:HDSSP}
\end{figure*}

In Figure \ref{fig:HDSSP}, we show an example of three such
iterations, converging rapidly to the real $\chi \left( x\right)
$, and in the inset a comparison with the $\chi \left( x\right) $
observed in simulations. Near the termination end of the system,
the EPD from simulations shows a depletion below the bulk density.
This reduction is characteristic of the high density phase and was
originally observed in the numerical results of \cite{MG}. The
continuum limit does not capture this feature.

A similar analysis may be carried out for the low density phase
when $\chi <\chi _{<}$. In this case, the boundary layer occurs at
the termination end, so it is convenient to apply the $\chi \left(
1\right) $ boundary condition explicitly and use the $\chi \left(
0\right) $ boundary condition to determine $J^{\ast }$. This
method also works for high and low density phase profiles when
$\chi \left( 0\right) ,\chi \left( 1\right) \in \left( \chi
_{<},\chi _{>}\right) $, as long as $\alpha $ and $\beta $ are not
too similar. The analysis fails close to the first order
transition line, as we would expect since Eqn.\ (\ref{SSPequ})
predicts a ``kink'' rather than a linear density profile on the
transition line.

For the maximal current phase, we know that $J^{\ast}>\hat{J}$, so
that $R$ is purely imaginary. Since $\chi ^{\prime }$ is negative
definite, we have a downward sloping profile, which is the
generalization of the $\tan \left( -x/\xi \right) $ (or $\cot
\left( x/\xi \right) $) profiles in the ordinary driven lattice
gas \cite{BSZ}. In the large $N$ limit, a macroscopically large
region of the profile will be almost flat (corresponding to $\chi
^{\prime }\simeq 0$), where the density assumes the optimal value
$\hat{\rho}$ (or $\chi \simeq \hat{\chi}$). Meanwhile, $J^{\ast}$
will approach the maximal current $ \hat{J}$ from above with
$O\left( 1/N^2\right) $ terms. The details are somewhat involved,
so that we will show the solution and discuss its properties only
for large $N$.

Defining the real quantity $\tilde{R}=-iR$, Eqn.\ (\ref{SSPequ})
can be integrated to show that the steady state density profile is
given by

\begin{widetext}
\begin{equation}
\ln \left[ \frac{\chi ^{2}-\left( 1-\tilde{J}\right) \chi
+J^{\ast}}{\chi _{0}^{2}-\left( 1-\tilde{J}\right) \chi
_{0}+J^{\ast}}\right] +\frac{4+2\left( \ell -1\right) \left(
1-\tilde{J}\right) }{\tilde{R}\left( \ell -1\right) }\left( \theta
-\theta _{0}\right) =-\frac{2x}{D \left( \ell -1\right) },
\label{implicitMC}
\end{equation}
\end{widetext}
where
\[
\theta \left( x\right) \equiv \arctan
\frac{\tilde{R}}{1-\tilde{J}-2\chi }
\]
and $\chi _{0}=\chi \left( 0\right) $ is given by Eqn.\
(\ref{chiBC0}). Evaluating Eqn.\ (\ref{implicitMC}) at the
termination end ($x=1,\chi =\chi \left( 1\right) $) and using
Eqn.\ (\ref{chiBC1}), we arrive at an equation for determining
$J^{\ast }$ in terms of the control parameters $\alpha ,\beta $.
Needless to say, this equation is too complex to solve
analytically. Nevertheless, we can gain some insight by
considering the large $N$ limit. Defining $\epsilon \equiv 1/N$
for convenience, recall that $D$ is $O\left( \epsilon \right) $.
Next, let us assume that $J^{\ast }=\hat{J}+O\left( \epsilon
^{2}\right) $ and show that it is justified later. This leads us
to $\tilde{R}$ being $O\left( \epsilon \right) $ and $\left(
1-\tilde{J}\right) /2=\hat{\chi}+O\left( \epsilon ^{2}\right) $,
so that $\chi _{>,<}=\hat{\chi}+O\left( \epsilon ^{2}\right) \pm
iO\left( \epsilon \right) $. Thus, the quadratic form in the
argument of the $\ln $ in Eqn.\ (\ref{implicitMC}) never becomes
smaller than $\tilde{R} ^{2}/4=O\left( \epsilon ^{2}\right) $, so
that this term never exceeds $ O\left( \ln N\right) $. Since the
other terms are generally $O\left( N\right) $ (from $D$ and
$\tilde{R}$), we can write an approximate equation by neglecting
the $\ln $ term, leaving an expression for $\theta ,$ and taking
the co-tangent of both sides:
\begin{equation}
\chi -\hat{\chi}=\frac{1}{2}\tilde{R}\cot \left[ x/\xi
+const.\right] +O\left( \epsilon ^{2}\right) \,\,,
\label{largeNMC}
\end{equation}
where $\xi \cong 2D\sqrt{\ell }/\tilde{R}$ is of $O\left( 1\right)
$. Note that the constant here can be written as
$\operatorname{arccot} \left[ 2\left( \chi _{0}-\hat{\chi}\right)
/\tilde{R}\right]$, which vanishes as $ \tilde{R}/\left( \chi
_{0}-\hat{\chi}\right) $ in the limit $\epsilon \rightarrow 0$.
Thus, for typical values of $x$, we see that $\chi =\hat{\chi
}+O\left( \epsilon \right) $. For the termination end, $J^{\ast }$
must be carefully fixed so that the argument of the co-tangent in
Eqn.\ (\ref {largeNMC}) approaches $\pi $ in an appropriate way
for the right hand side to equal $\chi \left( 1\right)
-\hat{\chi}$. This requirement is consistent with the original
assumption that $J^{\ast }=\hat{J}+O\left( \epsilon ^{2}\right) $.

\begin{figure}[tbp]
\centering
\includegraphics[clip=true]{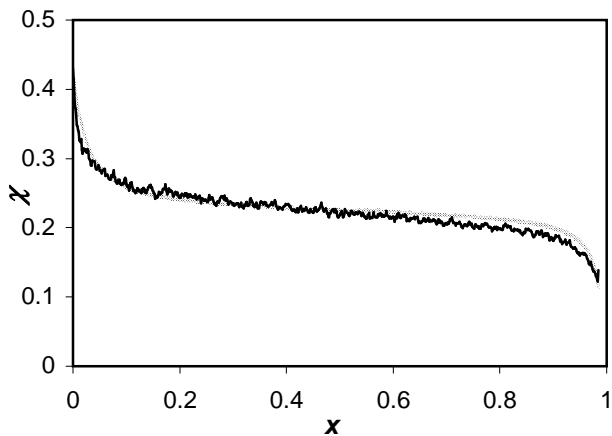}
\caption{Actual and predicted steady state density profiles in the
maximal current phase. The system considered here is $\alpha=0.5$,
$\beta=0.5$, $N=800$, $\ell=12$. $D$ was set to $4/5N$ to obtain a
good fit. $J^{\ast}$ was set to $\hat{J}+ 3.24 \times 10^{-5}$ to
satisfy the termination boundary condition. Profiles are the
actual steady state $\chi$ from simulations (bold curve) and the
predicted $\chi^{(7)}$ (lighter curve).} \label{fig:MCSSP}
\end{figure}

Returning to the finite $N$ case, although $\chi \left( x\right) $
cannot be determined explicitly, we find that an iterative method
is again helpful and converges rapidly. This time, successive
iterations of $\chi $ are substituted into the term $\ln \left[
\frac{\chi ^2-\left( 1-\tilde{J} \right) \chi +J^{\ast}}{\chi
_0^2-\left( 1-\tilde{J}\right) \chi _0+J^{\ast}}\right] $ of Eqn.\
(\ref{implicitMC}). As Figure \ref{fig:MCSSP} shows, there is good
agreement between the steady state profiles from simulation data
and this mean field theory.

\section{Disordered hopping rates in the open system}
\label{sec:disorder}

We finally turn to the difficult problem of the TASEP with
extended objects and quenched disorder. At each codon of the mRNA,
a ribosome translating the mRNA must wait for an appropriate
aa-tRNA to decode that codon. aa-tRNA availability ranges over
about an order of magnitude \cite{Solom}, and it is thought that
the elongation (hopping) rates may also range over an order of
magnitude, with the slowest ones comparable to the initiation and
termination rates \cite{Bergmann}.

Since previous studies of the TASEP with quenched disorder involve
$\ell =1$, their relevance to the process of translation may be
questioned. Indeed, we believe that the system will display
essential differences when these point particles are replaced by
extended objects. First, as we have seen in the uniform elongation
studies above, generalizing to the $\ell
>1$ case led to significant changes. Now, with non-uniform rates,
steric hinderance should play an even larger role, since extended
particles can block multiple sites with potentially different
rates. Thus, we will devote this section to a limited study of the
effects of $\ell >1$ on open systems with non-uniform elongation
rates: (a) systems with a single slow site, (b) bounds on the
current in general open systems, and (c) as an illustration of
systems with full disorder, simulation results for a real mRNA
sequence.

\subsection{Simple model of a single internal blockage}

We begin with the simplest form of quenched disorder, namely,
having a single \emph{internal} site $i$ with a reduced elongation
rate: $r<1$. If this blockage were moved to the termination site,
then $r$ would carry the label $\beta $ to conform with the
notation above. In most of the simulation studies for this
section, we focus on $\alpha =\beta =1$ with the blockage at the
center of the mRNA (i.e., $i=100$ with $N=200$). Though the
na\"{i}ve expectation is that the blockage should limit the
current in the same way, regardless of its location, we observe
that the current is noticeably reduced if the slow site is
internal. This reduction can be understood as follows. Once a
particle moves past the slow site, the particle behind it is not
necessarily free to move to the slow site. Instead, because of the
random sequential updating, the particle behind has a nonzero
chance of being blocked by the particle ahead. With just two
particles in the system, this effect can be accounted for exactly.
Beyond the scope of this paper, the details will be published
elsewhere. Here, let us present some simple heuristic arguments
which lead us to results that agree well with simulation data.

Extrapolating from the current in the uniform system (Eqn.\
\ref{Jab}), we speculate that
\begin{equation}
J=\rho _{source}\frac{P_{free}}T,  \label{Jslowint}
\end{equation}
where $\rho _{source}$ is the average coverage density behind the
slow site and $P_{free}$ is the approximate probability for a
particle just beyond the slow site to advance. Lastly, $T$ is the
average time to travel through the slow site and the $\ell -1$
sites that precede it. For example, in the termination-limited
case, $\rho_{source}=1-\beta$, $T=\ell-1+1/\beta$, and
$P_{free}=1$ (no steric hindrance beyond the termination site),
giving $J=\beta\left(1-\beta\right)/
\left[1+\beta\left(\ell-1\right)\right]$. We estimate $\rho
_{source}$ from the bulk density induced if the lattice were
truncated immediately after the slow site. $P_{free}$ can be
estimated from the density-dependent head spacing in a closed
system (Eqn.\ \ref{headspace}), using the bulk density that would
be induced if the lattice began with the slow site.

To determine $T$ approximately, we consider the behavior of two
particles in an infinite system with a single slow site (located
at the origin). Denoting the position of the leading and following
particles by $\zeta $ and $\eta $, respectively, the evolution of
this system can be regarded as a random walk in the $\zeta $-$\eta
$ plane, confined to $\zeta -\eta \geq \ell $. During each time
step, the walker moves, with equal probability, either ``upwards''
($\eta \rightarrow \eta +1$) or ``to the right'' ($\zeta
\rightarrow \zeta +1$). When it arrives at the $\zeta =\eta +\ell
$ line, the walker remains stationary with probability
$\frac{1}{2}$ (and moves ``to the right'' otherwise). Assuming
that there are $m+1$ holes between the particles immediately after
the leading one leaves the slow site, so that the walker is
``initially'' located at $\left( \zeta ,\eta \right) =\left(
1,-\ell -m\right) $, we computed numerically $\tau _{m}$, the
average time for the walker to arrive at the $\eta =0$ line (i.e.,
for the second particle to reach the slow site). Note that these
$\tau $'s take into account the steric hindrance due to the
leading particle, which we assume is always free to move. To
extract $T$, we make the assumption that the left particle
advances its first $m+1$ steps in time $m+1$, leaving $\ell -1$
more steps to reach the slow site. Now, the probability $ p\left(
m\right) $ for finding a gap of $m$ holes before the leading
particle passes the slow site can be estimated by inserting the
bulk density before the slow site ($\rho _{source}$) in Eqn.\
(\ref{headspace}). Taking the average over these initial starting
positions and accounting for the time to move over the slow site
($1/r$), we obtain $T=\sum_{m} \left[ \tau _{m}-\left( m+1\right)
\right] p\left( m\right) +1/r$. Predictions for the current from
Eqn.\ (\ref{Jslowint}), compared with the actual current from
Monte Carlo simulations, are shown in Table \ref {tab:currents}
for several values of the slow rate $r$. Despite the
approximations involved in this approach, the agreement is
surprisingly good.

\begin{table}
\caption{\label{tab:currents}Actual and predicted reductions in
current due to an internal slow site with rate $r$. Currents with
the slow site at one end are listed for comparison. All currents
are multiplied by $100$. Simulations were performed with
$N=200,\alpha =\beta =1$ with the blockage at the center. These
results are statistically the same if the blockage is placed at
site $20$ or $180$. }
\begin{ruledtabular}
\begin{tabular}{ccccc}
$r$ & $T_{\max }$ & $J$ for slow & actual $J$ & predicted $J$ \\
 & & site at end & & \\
\hline
0.01 & 13.5 & 0.89 & 0.86 & 0.87 \\
0.1 & 12.9 & 4.29 & 3.87 & 3.90 \\
0.2 & 12.3 & 5.00 & 4.68 & 4.57
\end{tabular}
\end{ruledtabular}
\end{table}

\subsection{Bounds for the current}

To model a real mRNA, we must allow for arbitrary translation
rates associated with each codon. Let us denote the rate at codon
$i$ by $k_i$. Due to the excluded volume constraint, however, it
is meaningful to consider also $K_{\ell ,i}$, the \emph{maximum}
rate for a ribosome to translate a stretch of $\ell $ sites
beginning with site $i$:
\[
K_{\ell ,i}\equiv \left( \sum_{q=i}^{i+\ell -1}\frac 1{k_q}\right)
^{-1}.
\]
Now, consider a ``window'' of any stretch of $\ell $ consecutive
sites in the lattice. If one particle is moving within this
window, the following particle must wait until the first one
passes through entirely before it can begin translating these
$\ell $ sites. Thus the characteristic time associated with the
current must be at least the time required to translate the
slowest stretch of $\ell $ codons. In this way, we find an upper
bound to the current, i.e.,
\begin{equation}
J \leq  \min_ {i\in \left\{ 1,\ldots ,N-\ell+1 \right\}}
 K_{\ell ,i} \, . \label{gatekeeper}
\end{equation}
One might imagine the slowest segment of $\ell $ codons acting as
a ``gatekeeper'' and preventing the current from exceeding the
value in Eqn.\ (\ref{gatekeeper}).

To arrive at a lower bound, we need only replace the elongation
rates at each site by the slowest elongation rate. From Eqn.\
(\ref{Jab}) above, we have the current of a system with uniform
rate unity. Thus, the minimum current for the disordered system is
simply
\begin{equation}
J\geq \left( \min_{i\in \left\{ 1,\ldots ,N-1\right\}} k_i\right)
\bar{J}\left( \alpha ,\beta \right) , \label{Jlb}
\end{equation}
where $\alpha$ is the ratio of the initiation rate to the slowest
elongation rate, and similarly for $\beta$. Though the gap between
these bounds for a real system may be too large to be of
significant predictive value, they can provide some guide to our
understanding of the current.

\subsection{Simulation of a real gene sequence}

To illustrate the full problem of disorder, we have simulated
translation of several real mRNA sequences from
\textit{Escherichia coli} strain MG1655, obtained from
\cite{MG1655}. Elongation rates at each codon were estimated using
commonly accepted values for the availability of tRNA in
\textit{E. coli} \cite{Solom}. The rate at each codon was assumed
proportional (with an arbitrary proportionality constant) to the
availability of the tRNA decoding that codon, as in \cite{Lesnik}.
Corresponding data were not available for estimating initiation
and termination rates, so a range of initiation and termination
rates was studied. We assumed that ribosomes cover $\ell =12$
codons \cite{Kang, Heinrich}.
\begin{figure}[tbp]
\centering
\includegraphics[clip=true]{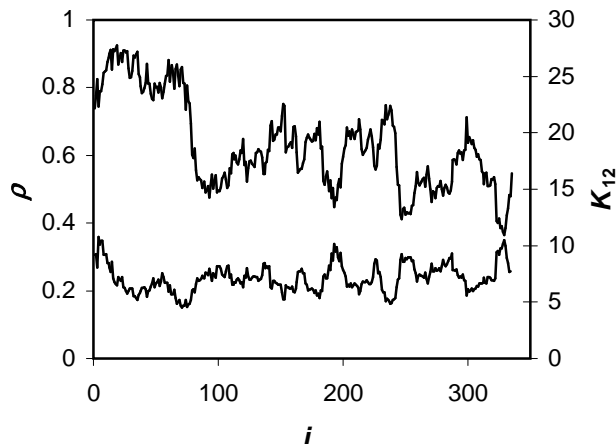}
\caption{Steady state coverage density $\rho$ (upper curve) and
maximum translation rate $K_{12}$ in each window of $\ell =12$ codons
(lower curve) for the \textit{ompA} gene of \textit{E. coli} when
the elongation rates are limiting. Elongation rates at each codon
were assumed proportional to availabilities of corresponding tRNA.}
\label{fig:realgene}
\end{figure}

Figure \ref{fig:realgene} shows the steady state coverage density
profile for the reasonably well studied gene \textit{ompA} when
the elongation rates are limiting. The lower curve in the figure
shows the maximum translation rate $K_{12}$ for each window of
$\ell =12$ codons, calculated from
\begin{equation*}
K_{12}=\left( \sum_{q=i}^{i+11}\frac{1}{k_q}\right) ^{-1}
\end{equation*}
for each site $i$. The minimum of the $K_{12}$ values is 4.51 in
arbitrary units, thus giving an upper bound for the current (Eqn.\
\ref{gatekeeper}). This value is significantly higher than the
actual current of 3.52. (For comparison, Eqn.\ \ref{Jlb} gives a
lower bound of 0.99 for the current in this system.) It should be
noted that the minimum $K_{12}$ occurs at codon 71, which is
approximately the location behind which the ribosome density is
very high, due to ribosomes ``piled up'' behind the slow region.
In general, lower values for the rate $K_{12}$ correspond to
higher ribosome densities, and higher $K_{12}$ to lower ribosome
densities, leading to an approximate symmetry between $K_{12}$ and
$\rho$. Thus the $K_{12}$ values are useful in understanding the
ribosome density profiles observed.

\section{Conclusions}
\label{sec:conclusions}

This work generalizes the well-studied $\ell =1$ TASEP model to
particles with extended sizes. Since there is a difference between
particle density $ \rho _r$ and density of occupied sites $\rho $,
the familiar particle-hole symmetry ($\rho \Leftrightarrow 1-\rho
$) takes the form $\rho _r\Leftrightarrow 1-\rho $ here. Exact
results for TASEP on a uniform ring, including probability
distributions for the current and for particle headway, were
found. Particularly useful in the latter part of this study is the
new current-density relationship, which is interpreted as a novel
density dependent mobility factor (or ``diffusion constant'').

An extremal principle \cite{PS} based on domain wall theory
allowed the closed system current-density relation to
predict currents and bulk densities in the uniform open system as
functions of the initiation and termination rates $\alpha$ and $\beta$.
The phase diagram for the open system was thus determined
using the extremal principle.
Domain wall theory also provided an explanation for the linear density
profiles and other unique characteristics observed at the first
order transition between the high and low density phases.
Given the ability of this theory to describe particles with
length $\ell >1$, it might be exploited further to determine
fluctuations in number of bound particles in the steady state
and behavior in the pre-steady state regime, as has been done
for the $\ell =1$ system \cite{SA}.

Based on the new mobility factor, a simple continuum limit led to
a differential equation for the density profile in open systems.
Though non-linear in general, this equation can be transformed, in
the case of a \textit{symmetric} exclusion process, to the
familiar \emph{linear} diffusion equation for an effective
particle density $\chi $. As a result, the stationary density
profile, expressed in terms of $\chi $, is again linear. For the
\textit{asymmetric} exclusion process, the differential equation
is more complex, due to an extra term to account for the drive.
Analytic expressions, albeit implicit, for the stationary profiles
were obtained and solved numerically through a rapidly converging,
iterative procedure. The results matched simulation data closely.
However, the predictive power of this approach is limited, since
the ``coarse-grained'' parameter $D$ cannot be derived from the
microscopic jump rates. Na\"{i}ve estimates for it are of the
right order of magnitude but quantitatively inaccurate, reflecting
the importance of particle correlations. Regarding $D$ as a
phenomenological quantity, we simply fit it to data. At a more
detailed level, open questions about density profiles remain. In
particular, our continuum limit cannot explain the particle
depletion and period-$\ell $ structure near the termination end of
high density systems. Profiles of the particle density $\rho _r$
are also of interest, though beyond the scope of this simple
continuum theory. Absent from the $\ell =1$ TASEP, peaks with
spacing $\ell $ extend far behind a blockage, with a decay length
well beyond typical microscopic scales (data not shown).
Evidently, when extended objects are included, even a simple
uniform TASEP tantalizes us with a rich variety of novel behavior.

Finally, systems with quenched disorder in the particle hopping
rates were briefly considered. Effects of a single internal slow
site on the current were estimated with fair accuracy by
considering the average delay a particle near the slow site
experiences due to a particle ahead of it. Also, bounds on the
current were determined for general disordered systems. A real
gene sequence was simulated, leading to a complicated density
profile. The parameter $K_{\ell ,i}$, the maximum rate to
translate a stretch of $\ell $ sites beginning with site $i$,
proved helpful in understanding the shape of the density profile,
but the disordered system remains far from solved.

We close with some speculation about the relevance of this work to
an understanding of translation. The advent of functional genomics
technologies to measure simultaneously mRNA and protein expression
profiles from many thousands of genes provides special
opportunities to begin to understand gene expression regulation.
Our results for currents (i.e., protein production rates) and
ribosome densities for uniform systems cannot be directly compared
with such experimental data from typical bacterial cells because
translation is not approximated well enough by a uniform system.
However, it is possible to use the uniform system results to
interpret data from an mRNA artificially constructed to be
uniform. Although there are no reports in the literature of
systems that are approximately uniform, the use of an \textit{in
vitro} translation system \cite{Cen} provides an opportunity to
make such experimental observations.

It is known that the relationship between mRNA and protein levels
in typical cells is nonlinear \cite{Gygi, Ideker}. Specifically,
when cells are grown under two different sets of conditions, the
amount of protein corresponding to a particular gene may be
down-regulated while its corresponding mRNA is up-regulated, and
the opposite may be true for other genes measured from the same
samples. We expect that in biological systems, the initiation rate
$\alpha$ should be an increasing function of the availability of
ribosomes within the cell. The protein production rate (current
$J$, Eqn.\ \ref{Jab}) is, in turn, a non-decreasing function of
$\alpha$. This analysis thus suggests that the observed nonlinear
relationship can arise from changes in the availability of
ribosomes given the nonlinear relationship between $J$ and
$\alpha$. However, this situation would cause all protein
production rates to change in the same direction, that in which
the ribosome availability changes. It would not permit some
proteins to be up-regulated while others are
down-regulated---which is the situation observed experimentally.
Further nonlinearity would arise if mRNA's were to compete for
available aa-tRNA as well as for ribosomes. Detailed modeling of
this effect will require a better understanding of systems with
quenched disorder, in which the elongation rates result from
aa-tRNA availability.

\begin{acknowledgments}
We thank Johannes Hager, James Sethna, Beate Schmittmann, Vassily
Hatzimanikatis, Amit Mehra, Tom Chou, and Anatoly Kolomeisky for
their helpful suggestions. RKPZ acknowledges support from the
National Science Foundation through grants DMR-9727574 and
0088451. KHL acknowledges support from the NSF through grants
BES-0120315 and 9874938. LBS was supported by an NSF Graduate
Research Fellowship and a Corning Foundation Fellowship. This
research was conducted using the resources of the Cornell Theory
Center, which receives funding from Cornell University, New York
State, federal agencies, foundations, and corporate partners.
\end{acknowledgments}

\appendix*
\section{Derivation of modified diffusion equation from discrete
version}

A modified diffusion equation qualitatively equivalent to Eqn.\
(\ref{SSPequ}) can be obtained by taking the na\"{i}ve continuum
limit of the discrete mean field equations of MacDonald \textit{et
al}. We begin with Eqn.\ (7) of \cite{MGP} for the current from
lattice site $j$ to $j+1$:
\begin{equation*}
q_{j} = \frac{n_{j} ^{\left( L \right) } n_{j+1} ^{\left( 0\right)
} } { n_{j+1} ^{\left( 0\right) } + n_{j+L} ^{\left( L \right) } }
\end{equation*}
and make the following correspondences with our continuum
notation:
\begin{eqnarray*}
n_j^{(0)} &\rightarrow &\rho _h\left( x\right) \\
n_j^{(L)} &\rightarrow &\rho _r\left( x\right) \\
L &\rightarrow &\ell \\
q &\rightarrow &J
\end{eqnarray*}
where $x$ lies in $\left[ 0,N\right] $ here. Thus we have
\begin{equation}
J \left( j \rightarrow j+1\right) =\frac{\rho _r\left( x\right)
\rho _h\left( x+1\right) }{\rho _h\left( x+1\right) +\rho _r\left(
x+\ell \right) } \label{MGPcont}
\end{equation}
Performing a series expansion of Eqn.\ (\ref{MGPcont}) and keeping
terms up to second order, we find that
\begin{widetext}
\begin{eqnarray*}
J \left( j \rightarrow j+1 \right) &=&\frac{\rho _r\left( x\right)
\left[ \rho _h\left( x\right) +\left( 1\right) \frac{\partial \rho
_h\left( x\right) }{\partial x}+\frac 1{2!} \frac{\partial ^2\rho
_h\left( x\right) }{\partial x^2}+...\right] }{\left[ \rho
_h\left( x\right) +\left( 1\right) \frac{\partial \rho _h\left(
x\right) }{\partial x}+\frac 1{2!}\frac{\partial ^2\rho _h\left(
x\right) }{
\partial x^2}+...\right] +\left[ \rho _r\left( x\right) +\ell \frac{\partial
\rho _r\left( x\right) }{\partial x}+\frac{\ell
^2}{2!}\frac{\partial ^2\rho _r\left( x\right) }{\partial
x^2}+...\right] } \\
&=& \frac{\rho _r\rho _h}{\rho _s}+\frac{\rho _r\rho _h^{\prime
}}{\rho _s} +\frac 12\frac{\rho _r}{\rho _s^2}\left[ \rho _r\rho
_h^{\prime \prime }-\ell ^2\rho _h\rho _r^{\prime \prime }\right]
\end{eqnarray*}
by using $\rho _h\left( x\right) =1-\ell \rho _r\left( x\right)$.
Similarly,
\begin{equation*}
J\left( j-1\rightarrow j\right) =\frac{\rho _r\rho _h}{\rho
_s}-\frac{\rho _r^{\prime }\rho _h}{\rho _s}+ \frac{\rho _r\rho
_h}{\rho _s^2}\rho _s^{\prime }+\frac{\rho _r\rho _h}{ 2\rho
_s}\left[ 2\left( \frac{\rho _s^{\prime }}{\rho _s}\right)
^2-2\frac{ \rho _r^{\prime }}{\rho _r}\frac{\rho _s^{\prime
}}{\rho _s}+\frac{\rho _r^{\prime \prime }}{\rho _r}-\left( \ell
-1\right) ^2\frac{\rho _r^{\prime \prime }}{\rho _s}\right]
\end{equation*}
\end{widetext}

It can then be shown that
\begin{eqnarray*}
\frac{\partial J}{\partial x} &=& J \left( j \rightarrow j+1
\right) - J\left( j-1\rightarrow j\right) \\
&=& \frac{\partial }{\partial x}\left[ \frac{ -1}{2 \rho_{s}
^{2}}\left[ \ell \left( 1-\ell \right) \rho_{r}^{2}+1\right]
\frac{\partial \rho_{r} }{\partial x}\right] + \frac{\partial
}{\partial x}\left( \frac{\rho _{r}\rho _{h}}{\rho _{s}}\right)
\,.
\end{eqnarray*}
This leads to the modified diffusion equation
\begin{equation}
J = D\left( \frac{1+\ell \left( 1-\ell \right)
\rho_{r}^{2}}{\rho_{s}^{2}}\right) \left( -\frac{\partial
\rho_{r}}{\partial x}\right) +E\frac{\rho _{r}\rho _{h}}{\rho
_{s}} \label{MGPSSP} \,,
\end{equation}
where we have defined $D=1/2$ and $E=1$.

Finally, we rewrite Eqn.\ (\ref{MGPSSP}) in terms of the effective
particle density $\chi$. Thus we find that the mean field
equations of \cite{MGP} predict
\begin{equation}
\frac{\partial \chi}{\partial x} = - \frac{1+\left( \ell -1
\right) \chi} {1+\left( \ell -1 \right) \chi -\left(\ell-1\right)
\chi^{2}} \left[ \tilde{J}\chi +J- \chi\left( 1-\chi \right)
\right] \label{MGPSSPchi} \,.
\end{equation}
We expect Eqn.\ (\ref{MGPSSPchi}) to be comparable to Eqn.\
(\ref{SSPequ}), the steady state density profile equation derived
previously from simple arguments. Indeed, both equations give the
same fixed points for $\chi$, thus producing qualitatively
identical density profiles. Further, their quantitative
differences have little effect on the shape of the density profile
(data not shown).


\begin{thebibliography}{43}
\bibitem{entropynote} If a ``coarse grained cell'' of $n$ sites
contains $n_{r}$ ribosomes and $ n_{h}$ holes, then the total
number of possible arrangements is $\left( n_{r}+n_{r}\right) !/$
$n_{r}!n_{h}!$. Writing $n_{r}=n\rho _{r}$, etc., and taking the
logarithm leads to the expression here.
\bibitem{linearnote} Note that it is always possible to achieve a
linear profile (in these one-dimensional systems) by an
appropriate transformation of the density variable, as long as the
effective ``diffusion constant'' is a single valued function of
$\rho ^{*}$.
\bibitem{inflimitnote} In this section, we will use only the the
$N\rightarrow \infty $ limit of Eqn. (\ref{Jvsrho}).
\end{thebibliography}
\end{document}